\documentclass{emulateapj}
\usepackage{graphicx}
\usepackage{amssymb}
\usepackage{natbib}

\newcommand{\ha}{H$\alpha$}
\newcommand{\hb}{H$\beta$}

\def\ltsima{$\buildrel<\over\sim$}
\def\lsim{\lower.5ex\hbox{\ltsima}~}
\def\gtsima{$\buildrel>\over\sim$}
\def\gsim{\lower.5ex\hbox{\gtsima}~}
\def\msol{M$_{\odot}$}

\def\msolyr{M$_{\odot}$~yr$^{-1}$}

\def\teff{\ifmmode T_{\rm eff} \else $T_{\mathrm{eff}}$\fi}

\def\lya{Ly$\alpha$} 
\def\ha{H$\alpha$} 
\def\hb{H$\beta$}

\def\ergscm{erg~s$^{-1}$~cm$^{-2}$}
\def\cm2{cm$^{-2}$}

\def\hii{H{\sc ii}}

\def\oiii{O{\sc iii}}
\def\oii{O{\sc ii}}
\def\nii{N{\sc ii}}
\def\sii{S{\sc ii}}
\def\nh{\ifmmode N_{\mathrm{HI}}\else $N_{\mathrm{HI}}$\fi}

\def\izw{{\sc I}Zw 18}
\def\vexp{\ifmmode v_{\rm exp} \else v$_{\rm exp}$\fi}
\def\taua{\ifmmode \tau_{a}\else $\tau_{a}$\fi}

\newcommand{\grisma}{$G_{102}$}
\newcommand{\grismb}{$G_{141}$}

\begin{document} 


\title{Very Strong Emission-Line Galaxies in the WISP Survey And Implications For High-Redshift Galaxies \footnote{Based on observations made with the NASA/ESA Hubble Space Telescope, which is operated by the Association of Universities for Research in Astronomy, Inc., under NASA contract NAS 5-26555. These observations are associated with programs 11696 and 12283.} \footnote{Some of the data presented herein were obtained at the W.M. Keck Observatory, which is operated as a scientific partnership among the California Institute of Technology, the University of California and the National Aeronautics and Space Administration. The Observatory was made possible by the generous financial support of the W.M. Keck Foundation} 
}

\author{Atek H. \altaffilmark{1},
Siana B. \altaffilmark{2},
Scarlata C.\altaffilmark{3},
Malkan M.\altaffilmark{4},
McCarthy P.\altaffilmark{5},
Teplitz H.\altaffilmark{6},\\
Henry A.\altaffilmark{7},
Colbert J.\altaffilmark{1},
Bridge, C.\altaffilmark{2},
Bunker, A.J.\altaffilmark{8},
Dressler, A. \altaffilmark{5},\\
Fosbury, R.A.E. \altaffilmark{9},
Hathi, N. P. \altaffilmark{5},
Martin, C. \altaffilmark{7},
Ross N. R. \altaffilmark{4},
Shim, H. \altaffilmark{1}}

\altaffiltext{1}{Spitzer Science Center, Caltech, Pasadena, CA 91125}
\altaffiltext{2}{Department of Astronomy, Caltech, Pasadena, CA 91125}
\altaffiltext{3} {University of Minnesota - Twin Cities, Minneapolis, MN 55455}
\altaffiltext{4}{Dep't. of Physics and Astronomy, Univ. of Calif. Los Angeles}
\altaffiltext{5}{Observatories of the Carnegie Institution for Science, Pasadena, CA 91101} 
\altaffiltext{6}{Infrared Processing and Analysis Center, Caltech, Pasadena, CA 91125}
\altaffiltext{7}{Dep't. of Physics, Univ. of Calif. Santa Barbara, CA 93106}
\altaffiltext{8}{Department of Physics, University of Oxford, Denys Wilkinson Building, Keble Road, OX13RH, U.K.}   
\altaffiltext{9}{Space Telescope - European Coordinating Facility, Garching bei M\"unchen, Germany}

\shorttitle{A Population of High-EW Emission-Line Galaxies in the WISP Survey}
\shortauthors{Atek et al}

\begin{abstract}

The WFC3 Infrared Spectroscopic Parallel Survey (WISP) uses the {\em Hubble Space Telescope} infrared grism capabilities to obtain slitless spectra of thousands of galaxies over a wide redshift range including the peak of star formation history of the Universe. We select a population of very strong emission-line galaxies with rest-frame equivalent widths higher than 200 \AA. A total of 176 objects are found over the redshift range $0.35 < z < 2.3$ in the 180 arcmin$^2$ area that we have analyzed so far. This population consists of young and low-mass starbursts with high specific star formation rates (sSFR). After spectroscopic follow-up of one of these galaxies with {\em Keck}/LRIS, we report the detection at $z = 0.7$ of an extremely metal-poor galaxy with 12+Log(O/H)$= 7.47 \pm 0.11$. After estimating the AGN fraction in the sample, we show that the high-EW galaxies have higher sSFR than normal star-forming galaxies at any redshift. We find that the nebular emission-lines can substantially affect the total broadband flux density with a median brightening of 0.3 mag, with some examples of line contamination producing brightening of up to 1 mag. We show that the presence of strong emission lines in low-$z$ galaxies can mimic the color-selection criteria used in the $z \sim 8$ dropout surveys. In order to effectively remove low redshift interlopers, deep optical imaging is needed, at least one magnitude deeper than the bands in which the objects are detected. Without deep optical data, most of the interlopers cannot be ruled out in the wide shallow {\em Hubble Space Telescope} imaging surveys. 
Finally, we empirically demonstrate that strong nebular lines can lead to an overestimation of the mass and the age of galaxies derived from fitting of their SED. Without removing emission lines, the age and the stellar mass estimates are overestimated by a factor of 2 on average and up to a factor of 10 for the high-EW galaxies. Therefore the contribution of emission lines should be systematically taken into account in SED fitting of star-forming galaxies at all redshifts. 
\end{abstract}
\maketitle

\section{Introduction}\label{sec:intro}

The characterization of galaxies undergoing their first major star formation episode is of great interest in modern astrophysics. This class of galaxies is expected to host young and massive stars ionizing the interstellar medium, and therefore to exhibit strong nebular emission lines. However, many efforts dedicated to the understanding of galaxy formation and evolution are in general limited by different factors or colored by the way in which galaxy samples are assembled. Broadband surveys yield magnitude-limited samples that are biased towards bright continuum objects such as the Lyman Break Galaxies \citep[LBGs,][]{steidel96,shapley03,vanzella09,hathi10} , and can favor relatively massive galaxy populations. On the other hand, ground-based searches for rest-frame optical emission lines from high-redshift galaxies are severely impacted by the bright NIR background. Observing in small wavelength windows between OH terrestrial airglow will significantly restrict the survey volume.

In this context, space-based observations offer considerable advantages for the detection of $z \sim 1-2$ emission-line galaxies. The Wide field Camera 3 (WFC3) onboard the {\em Hubble Space Telescope} ({\em HST}) has dramatically improved the power of slitless spectroscopy from above the atmosphere. Beginning in cycle 17, we are conducting the WFC3 Infrared Spectroscopic Parallel Survey (WISP). In about 500 orbits of parallel observations, the program will obtain slitless spectra of galaxies at the peak of the star formation history of the Universe \citep[e.g.][]{Hopkins04}. The \grisma\ and \grismb\ IR grisms offer continuous wavelength coverage from 0.8 to 1.7 \micron, allowing the selection of emission-line galaxies at $0.35 < z < 2.5$, and potentially bright \lya\ emitters at $z > 6.5$ \citep[see][for a complete description of the survey] {atek10}.

In this paper, we present a sample of emission-line objects with extremely high equivalent widths. The equivalent width of recombination lines indicates the ratio of the current star formation rate (SFR) to the past average SFR of a galaxy \citep[e.g.][]{kennicutt98}. Therefore, the high-EW selection picks up galaxies with a strong ongoing star formation episode. 

Previously, ground-based narrow-band surveys were used to identify a large sample of ultra strong emission lines galaxies \citep[USELs,][]{kakazu07,hu09}, but were restricted to a narrow range of redshifts up to $z = 1$. In the local universe, \citet{cardamone09} found in the Sloan Digital Sky Survey (SDSS) compact star-forming galaxies, called ``green peas'', that have very blue $r - i$ colors because of their very strong [\oiii]$\lambda$5007 emission line. Thanks to the unprecedented sensitivity of the WFC3/IR detectors, the depth of the direct imaging combined with the resolution and the wavelength coverage of the grism spectroscopy allows us to uncover this particular class of objects independently of the continuum brightness over a wide redshift range.

The selection of high-EW star-forming galaxies with the WFC3 grisms probes a lower mass range than previous studies at $z > 1$ \citep[e.g.][]{erb06b,daddi07}. This allows us to compare the star-formation efficiency of these dwarf galaxies with normal star-forming galaxies at $z \sim 1-2$.  Indeed, several studies have established a correlation between the star formation rate and the stellar mass up to redshift $z = 7$ \citep[e.g.][]{brinchmann00, elbaz07, noeske07, daddi07,pannella09,oliver10,labbe10}. The shape and evolution of this relationship is interpreted as the result of smooth gas accretion \citep{bouche10,dave11}. However, at high redshifts, the studies focused on a specific mass range due to the difficulty of selecting faint-continuum galaxies.

In a broader context, we also discuss the implications of these strong line-emitters on the selection of $z \sim 8$ galaxies and on the physical properties of galaxies derived from SED modeling.

We present in Section \ref{sec:obs} the WISP survey observations and the data reduction. Section \ref{sec:follow} is devoted to the follow-up observations. The selection of the high-EW population is presented in Section \ref{sec:high_ew}. In Section \ref{sec:ssfr} we discuss the mass and star-formation properties of the high-EW galaxies. We present in Section \ref{sec:low_metal} the metallicity measurement of one example and put it into context. We investigate the contribution of nebular lines to the total broadband flux density and their astrophysical implications in Section \ref{sec:neb_contrib} and \ref{sec:sed}. Our conclusions are given in Section \ref{sec:summary}. Throughout, we assume a
$\Lambda$-dominated flat universe, with $H_0=70$\ km s$^{-1}$\
Mpc$^{-1}$, $\Omega_{\Lambda}=0.7,$\ and $\Omega_{m}=0.3$. All magnitudes are in AB system.

\begin{figure*}[!htbp]
   \centering
   \hspace{-1cm}   
     \includegraphics[width=7.8cm]{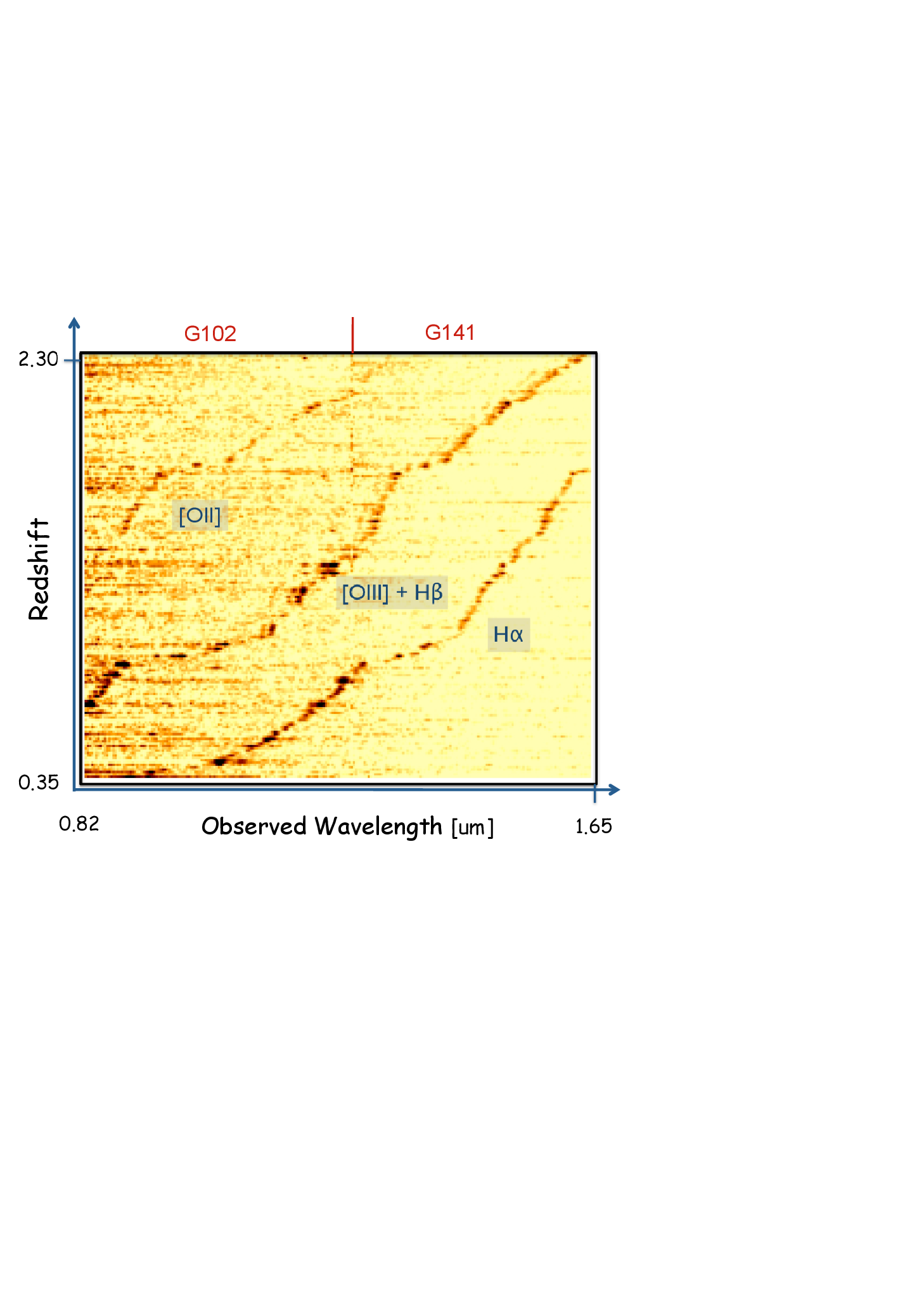}
    \hspace{-3cm}
   \includegraphics[width=10.5cm]{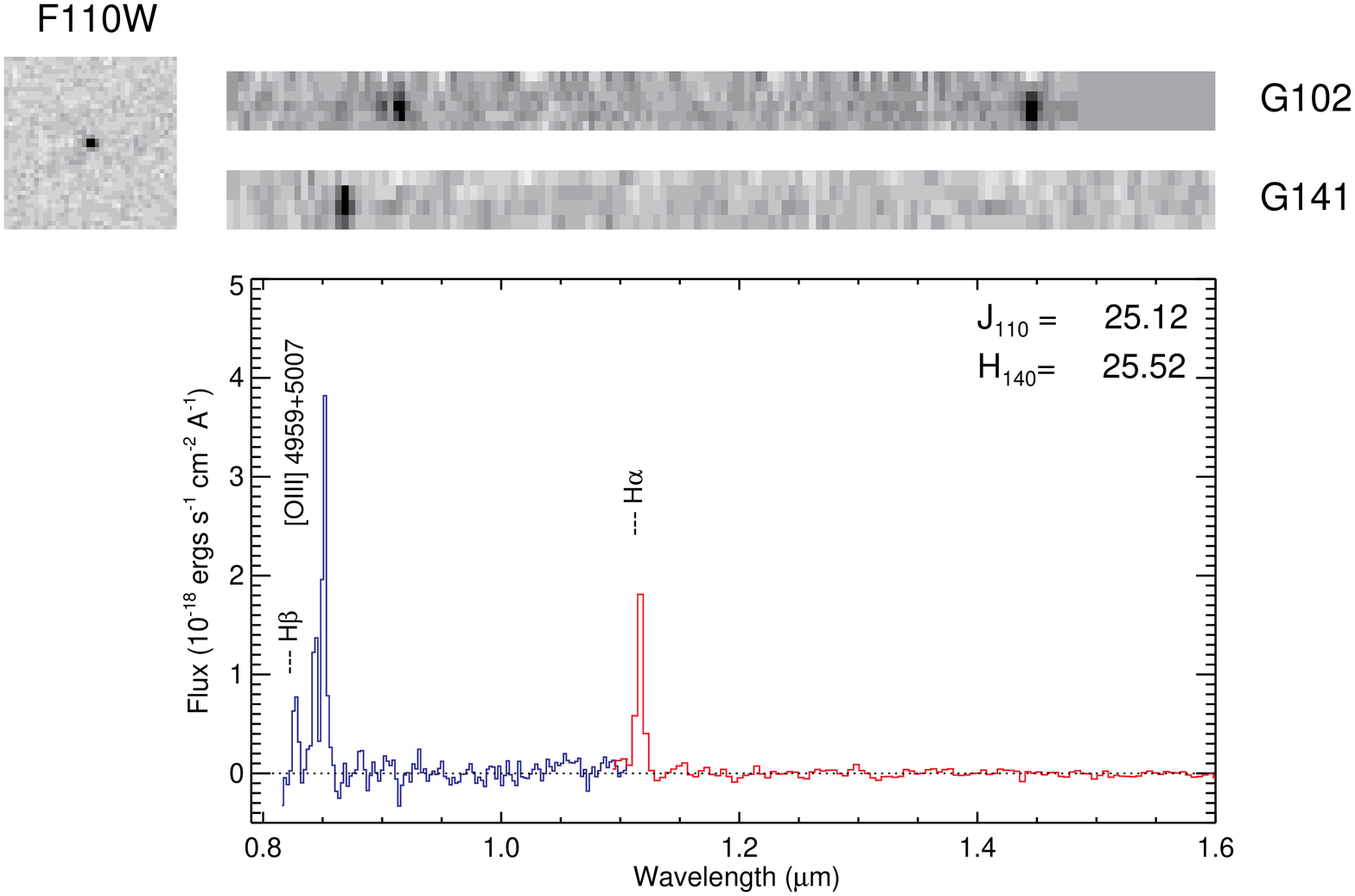} 
    \hspace{-3.5cm}
   \caption{Strong emission-line galaxies across a wide range of redshift in the WISP Survey. {\it Left:} Montage spectra of 176 high-EW emission-line galaxies stacked in the Y-axis direction with increasing redshift. Consequently, we see the emission lines we routinely detect in WISPS being shifted in the wavelength dimension. We note that the slope of the emission-line curve is less steep in some parts of this plot. i.e. we detect less objects per redshift bin, which is due the decreasing sensitivity towards the edge of the grisms. {\it Right:} An example of a galaxy at $z \sim 0.7$ showing strong emission lines. The direct F110W image and the G102 and G141 2D spectra are shown above the calibrated 1D spectrum. The $J$ and $H$ magnitudes are also given in the inset. The object shows bright \ha\ and [\oiii] lines at 0.85 and 1.12 $\mu$m, respectively, on top of a very faint continuum. The \ha\ line is detected in both grisms, in the overlapping region.}
   \label{fig:examples}
\end{figure*}

\section{The WISP Survey}
\label{sec:obs}

The WISP survey exploits the powerful pure parallel mode of the {\it HST} to obtain slitless spectra of nearly 200 uncorrelated fields in 500 orbits within two observing programs (PI = M. Malkan): GO 11696 and 12283 (Infrared Survey of Star Formation Across Cosmic Time). During long integrations with the Cosmic Origin Spectrograph \citep[COS,][]{osterman11} or the Space Telescope Imager and Spectrograph \citep[STIS,][]{woodgate98}, the WISP program uses WFC3 \citep{kimble08} to observe an adjacent field in parallel, at a fixed offset of about 5\arcmin\ from the primary target. The data include slitless spectroscopy with G102 (0.8 - 1.2 \micron) and G141 (1.1-1.7 \micron) grisms, and direct NIR imaging with the F110W and F140W filters. As of August 2010, we added the F475X and F600LP filters of the WFC3/UVIS channel to our observing program, with typical exposure times of 400 sec in each band. The fields analyzed in this paper do not have UVIS observations. For the long visits (typically 4 orbits or more), we observe with both grisms with a typical exposure time ratio of G102:G141 = 2.5:1. For shorter visits the fields are observed in G141 only. The direct images are obtained in the same orbits as the grism imaging with a $\sim$ 6:1 grism:direct integration time ratio. It is necessary for the spectral extraction from slitless data to use the direct images to provide the input catalog of objects. The object positions, sizes and shapes are then used, in conjunction with a dispersion solution that translates the position from the direct image to the spectral trace in the grism one, to extract flux and wavelength calibrated 2D and 1D spectra. 

In this paper, we analyzed a total of 54 fields presented in Table \ref{tab:obs}, among which two have optical follow-up (cf. Section \ref{sec:follow}). We note that our effective area represents about 70\% of the full WFC3/IR frame because of two things. First, objects that fall outside the right edge of the direct image frame will have their zero order spectra falling in the right side of the grism frame (zeroth order is dispersed to the left side of the object position). Thus, we cannot correct the zeroth order contamination in that part of the detector. Second, the left edge of the grism frame will receive spectra from objects that are outside the direct image (first order dispersed to the right side of the object position), preventing any flux or wavelength calibration for those sources. 

All the data were processed with the WFC3 pipeline CALWF3 (version 2.1) to correct for bias, dark, flatfield and gain variations. Then, the slitless extraction package aXe 2.0 \citep{kuemmel09} is used for the spectral extraction. Spectra with important contamination are removed from the analysis because the contamination estimate of aXe is not sufficiently accurate. A complete description of the data reduction steps is presented in \citet{atek10}.

\section{Optical Follow-Up Spectroscopy and Imaging}
\label{sec:follow}

We have obtained optical spectra of objects in two adjacent WISP fields with the Low Resolution Imaging Spectrometer \citep[LRIS,][]{oke95,steidel04} on the {\it Keck I} telescope. The two WFC3 fields, WISP5 and WISP7 in Table \ref{tab:obs}, are immediately adjacent to each other and fit within one LRIS slit mask.  We assigned 1.2$''$ wide slits to 13 emission line objects, 5 of which belong to the high-EW sample. Three exposures of 1800s each were used for a total integration of 5400s. The 400 l/mm grism blazed at 3400 \AA\ was used on the blue arm and the 600 l/mm grating blazed at 1.0 $\mu$m was used on the red arm for a pixel scale of $\sim 1.1$ \AA\ pix$^{-1}$ and  $\sim 0.8$ \AA\ pix$^{-1}$, respectively. The plate scale is 0.135\arcsec\ pix$^{-1}$ and the spectral resolution for an object that fills the slit is about 8.1 \AA\ and 5.6 \AA\ in the blue and red side, respectively. The seeing during observations was around 1\arcsec. The spectrophotometric standard Wolf 1346 was used to flux-calibrate the spectra.

\begin{deluxetable*}{l c c c c c c c c}
\tabletypesize{\small}
\tablecolumns{9}
\tablewidth{370pt}
\centering
\tablecaption{Summary of WFC3 Imaging and Spectroscopy \label{tab:obs}}
\tablehead{
\colhead{Field}  & \colhead{   RA} & \colhead{DEC} & \colhead{F110W} & \colhead{G102} & \colhead{F140W} & \colhead{G141}  \\
\colhead{ }        & \colhead{(HMS)}  & \colhead{(DMS)}  & \colhead{(sec)} & \colhead{(sec)} & \colhead{(sec)} & \colhead{(sec)}  
}
\startdata
       
WISP1  &  01 06 35.29  & $+$15 08 53.8  &  884   &   4815  &   506   &	2609   \\
WISP2  &  01 25 10.02  & $+$21 39 13.7  &  0     &   0	 &   328   &	1906   \\     
WISP5$^{\ast}$  &  14 27 06.64  & $+$57 51 36.2  &  1034  &   5515  &  1034   &	5515   \\
WISP6  &  01 50 17.18  & $+$13 04 12.8  &  609   &   3609  &   862   &	5015   \\
WISP7$^{\ast}$  &  14 27 05.91  & $+$57 53 33.7  &  834   &   6318  &  1112   &	6224    \\
WISP8  &  11 51 51.62  & $+$54 40 41.2  &  1662  &   9021  &   581   &	2509   \\
WISP9  &  12 29 44.31  & $+$07 48 23.5  &  759   &   4612  &   684   &	3712   \\
WISP10 &  09 25 07.84  & $+$48 57 03.0  &  631   &   3909  &   406   &	2209   \\
WISP11 &  11 02 17.38  & $+$10 54 25.4  &  556   &   3709  &   456   &	2006   \\
WISP12 &  12 09 25.25  & $+$45 43 19.8  &  1312  &   8221  &   606   &	3009   \\
WISP13 &  01 06 38.77  & $+$15 08 26.2  &  556   &   3009  &   506   &	2409   \\
WISP14 &  02 34 56.80  & $-$04 06 54.5  &  834   &   6215  &   481   &	2809   \\
WISP15 &  14 09 42.47  & $+$26 21 56.0  &  1612  &   8321  &   531   &	2609   \\
WISP16 &  02 34 54.72  & $-$04 06 42.5  &  1087  &   6921  &   584   &	 2509   \\
WISP18 &  12 29 17.25  & $+$10 44 00.6  &  534   &   3512  &   762   &	 3824   \\  
WISP19 &  02 34 54.29  & $-$04 06 30.5  &  1187  &   8721  &   484   &	 2809   \\  
WISP20 &  14 09 41.15  & $+$26 22 15.1  &  1815  &   8430  &   559   &	 2812   \\
WISP21 &  09 27 55.77  & $+$60 27 05.3  &  0     &   0	 &   353   &	 2006   \\
WISP22 &  08 52 44.99  & $+$03 09 09.6  &  0     &   0	 &   253   &	 1806   \\
WISP23 &  09 43 16.12  & $+$05 27 37.1  &  0     &   0	 &   681   &	 4115   \\
WISP24 &  12 18 41.90  & $+$29 52 51.8  &  0     &   0	 &   278   &	 1806   \\
WISP25 &  10 08 42.49  & $+$07 11 10.3  &  0     &   0	 &   734   &	 4115   \\
WISP26 &  08 45 17.91  & $+$22 54 58.5  &  840   &   5521  &   384   &	 2209   \\
WISP27 &  11 33 05.98  & $+$03 28 02.2  &  1040  &   6218  &   384   &	 2206   \\
WISP28 &  09 35 46.24  & $+$14 27 48.5  &  0     &   0	 &   634   &	 3515   \\
WISP29 &  12 02 59.55  & $+$48 05 21.2  &  0     &   0	 &   328   &	 1906   \\
WISP30 &  10 28 18.91  & $+$39 17 14.1  &  0     &   0	 &   253   &	 1806   \\
WISP31 &  08 43 27.99  & $+$26 16 39.7  &  0     &   0	 &   659   &	 4412   \\
WISP32 &  13 05 20.67  & $-$25 38 05.5  &  0     &   0	 &   484   &	 3512   \\  
WISP33 &  10 00 02.78  & $+$12 44 56.4  &  959   &   6415  &   484   &	 2809   \\  
WISP34 &  09 34 58.06  & $+$02 02 03.7  &  0     &   0	 &   709   &	 4215   \\  
WISP35 &  13 03 46.90  & $+$29 53 03.8  &  0     &   0	 &   634   &	 3812   \\
WISP36 &  13 40 31.52  & $+$41 23 10.3  &  1237  &   8621  &   584   &	 2809   \\
WISP38 &  12 25 13.33  & $-$02 49 08.4  &  0     &   0	 &   709   &	 4312   \\
WISP39 &  10 09 38.23  & $+$30 00 48.7  &  0     &   0	 &   328   &	 1706   \\
WISP40 &  02 16 20.12  & $-$39 02 28.1  &  0     &   0	 &   584   &	 4112   \\
WISP41 &  12 08 28.31  & $+$45 38 48.2  &  1084  &   6615  &   584   &	 2909   \\
WISP42 &  10 01 04.36  & $+$50 24 02.8  &  1812  &   10424 &   784   &	 4012   \\
WISP43 &  21 04 07.62  & $-$07 23 00.8  &  909   &   6315  &   434   &	 2809   \\
WISP44 &  11 12 15.18  & $+$35 36 55.5  &  0     &   0	 &   734   &	 4515   \\
WISP45 &  12 37 26.68  & $+$01 25 18.2  &  0     &   0	 &   796   &	 4812   \\
WISP46 &  22 37 58.48  & $-$18 42 05.4  &  0     &   0	 &   328   &	 1906   \\
WISP46 &  22 37 58.48  & $-$18 42 05.4  &  0     &   0	 &   328   &	 1906   \\
WISP47 &  13 19 32.40  & $+$27 26 40.0  &  0     &   0	 &   587   &	 3212   \\  
WISP49 &  14 44 44.84  & $+$34 27 45.9  &  1093  &   6018  &   406   &	 2406   \\  
WISP50 &  22 22 20.25  & $+$09 36 36.8  &  0     &   0	 &   328   &	 1906   \\
WISP51 &  15 13 13.77  & $+$36 33 22.0  &  0     &   0	 &   278   &	 2006   \\
WISP52 &  13 30 22.62  & $+$28 11 01.0  &  0     &   0	 &   634   &	 4218   \\
WISP53 &  15 11 12.91  & $+$40 25 39.3  &  0     &   0	 &   253   &	 1806   \\
WISP54 &  15 44 57.26  & $+$48 45 23.2  &  0     &   0	 &   303   &	 2006   \\
WISP56 &  16 16 50.58  & $+$06 36 43.6  &  0     &   0	 &   684   &	 4312   \\
WISP57 &  12 33 16.74  & $+$47 53 27.5  &  0     &   0	 &   659   &	 4615   \\
WISP59 &  15 50 22.59  & $+$39 59 19.2  &  0     &   0	 &   734   &	 4515   \\
WISP61 &  23 09 00.35  & $-$09 07 17.7  &  0     &  0	 &   709   &   4312

\enddata
\tablecomments{Observation information for the fields analyzed in this paper. WISP5 and WISP7 have optical imaging in $g'$ and $i'$ bands with exposure times of 1500 and 3000 sec, respectively. The two fields were also covered by a slitmask with LRIS at {\em Keck I} for a total exposure time of 5400 sec. 1.2\arcsec wide slits were used.}
\end{deluxetable*}

We also observed these two fields on April 20 2010 to obtain optical images with the Large Format Camera (LFC) on the {\em Hale} 5m telescope at Palomar observatory. The two fields were easily covered by the mosaic camera with a field of view of 24 arcmin diameter. We used the $g'$ and $i'$ filters with a total exposure time of 1500 and 3000 sec, respectively. On readout, we binned $2\times2$, resulting in a pixel scale of 0.36\arcsec\ per pixel. The typical seeing during observations was about 1\arcsec. We adopted a 5 point dithering pattern with a 15\arcsec\ displacement. A summary of the observations is given in Table \ref{tab:obs}.

The images were corrected for bias and flatfield using standard \texttt{IRAF} reduction packages. The flux calibration was performed by comparing the photometry of the field stars of our images with the those of the Sloan Digital Sky Survey \citep[SDSS][]{york00} that we obtained from the data release DR8. Finally, we aligned the ground-based frames to our IR ones using \texttt{GEOMAP} and \texttt{GEOTRAN} tasks in \texttt{IRAF}. In order to measure the object magnitudes in the optical images we need to match the photometry aperture between ground-based and the {\em HST} images. We first convolve the WFC3 images with a gaussian kernel to match the PSF of the Palomar data. Then, we measure the flux in the convolved images using an aperture of 5 WFC3/IR pixels. The aperture size adopted is large enough to encompass most of the object flux for our compact sources, without being contaminated by nearby sources. We then compare the measured magnitudes to the AUTO magnitudes measured with {\tt SExtractor} in the original WFC3 images. This gives us the aperture correction to apply to the magnitudes measured in the optical data.

\section{High Equivalent-Width Population}
\label{sec:high_ew}

The WISP survey is particularly sensitive to high specific SFR and strong emission-line galaxies. In the fields we observed so far, we often see isolated emission lines in the grism frames, with either a faint continuum or no continuum at all. Such objects are easily identifiable by visually inspecting the final 1D spectra. We only search part of the field, for which the zeroth order information is available. To this aim, we have developed a flagging scheme in our reduction pipeline that marks the position of the zeroth order in the 2D and 1D spectra, to prevent their identification as emission lines. We then use a custom IDL procedure based on a least square method to fit all the lines in each spectrum with a Gaussian and a polynomial continuum model. 

\begin{figure}[htbp]
\centering
\hspace{-0.7cm}
   \includegraphics[width=9cm]{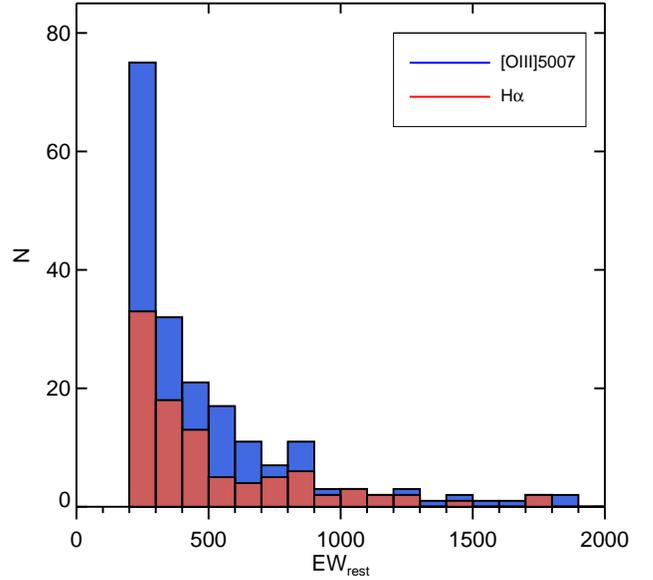} 
   \caption{Rest-frame equivalent width distribution for objects with EW $ \geq 200$ \AA\ in the WISP Survey. The total number in each bin is divided into the [\oiii] $\lambda$5007 line (presented in blue) and the \ha\ line (presented in red).}
   \label{fig:ew_histo}
\end{figure}

After the line measurement we have retained only galaxies with rest-frame EW higher than 200 \AA. Examples of strong emission-line objects in our catalog are presented in Figure \ref{fig:examples}. The strong lines are typically \ha\ and [\oiii]. The sample reaches a 3$\sigma$ flux limit of $7 \times 10^{-17}$ \ergscm\ over a 180 arcmin$^{2}$ area, which is about three times the area surveyed in \citet{atek10}. Both IR grisms are used to observe 24 fields (80 arcmin$^{2}$) over a wavelength range of 0.8-1.7 \micron, and 30 fields (100 arcmin$^{2}$) are observed with G141 only covering a wavelength range 1.1-1.7 \micron. In Figure \ref{fig:ew_histo}, we present the rest-frame equivalent width distribution of the high-EW sample. It is well known that the \ha\ equivalent width is an indicator of the age of a galaxy \citep[e.g.][]{leitherer99}. The \ha\ equivalent width indicates the ratio of the current SFR to the lifetime averaged SFR, also known as the birthrate parameter {\em b} \citep{scalo86,kennicutt98}. The continuum luminosity is representative of the low mass stars with a lifetime of several Gyrs. The strength of \ha\ line evolves on much shorter timescales, as it is the result of hot, short lived O and B stars.  A galaxy with a constant star formation will have a birthrate parameter of unity, while $b > 1$ indicates that the current episode of star formation is stronger than the average SFR.  Using {\tt Starburst99} models \citep{leitherer99}, we find that the 200 \AA\ limit on EW(\ha) selects either young galaxies (less than 10 Myr) with instantaneous burst episode, or galaxies with continuous star formation older than a burst and less than 1 Gyr.

We note that for extremely high-EW lines, the continuum remains undetected in the grism spectra, making the EW measurement highly uncertain. Given the low spectral resolution of the WFC3 grism we are not able to separate the [\nii] lines from \ha. However, because the high-EW galaxies are likely metal-deficient (see Section \ref{sec:low_metal}), we do not expect a large contribution from the [\nii] lines. In fact, given an [\oiii] $\lambda$5007/\hb\ ratio, one can infer the [\nii]/\ha\ ratio from the [\sii]/\ha\ one. From Figure 1 of \citet{kewley06}, and using the measured [\sii]/([\nii]+\ha), we derive the [\sii]/\ha\ and then [\nii]/\ha\ for 17 high-EW galaxies. We found a median contribution of [\nii] to \ha\ of about 8 \%.

The strong [\oiii] $\lambda$5007 line can also be the result of a photoionization dominated by harder radiation field from an Active Galactic Nucleus (AGN). This will favor collisionally-excited lines rather than recombination lines. In order to estimate the AGN contribution in the sample we use the BPT diagnostic diagram proposed by \citet{veilleux87} using [\oiii] $\lambda$5007/\hb\ versus [\sii] $\lambda$(6717+6732)/\ha\ line ratios, where \ha\ is corrected for [\nii] contamination following the procedure described above. We note that using the BPT diagram to classify low-metallicity dwarf galaxies with high sSFR can be somewhat ambiguous  \citep{brinchmann08, izotov08}. \citet{brinchmann08} show that high-sSFR galaxies may have higher values of the ionization parameter than normal star-forming galaxies, and tend to be offset in the BPT diagram towards the AGN region. 

\begin{figure}[htbp]
\centering
\hspace{-0.7cm}
   \includegraphics[width=9cm]{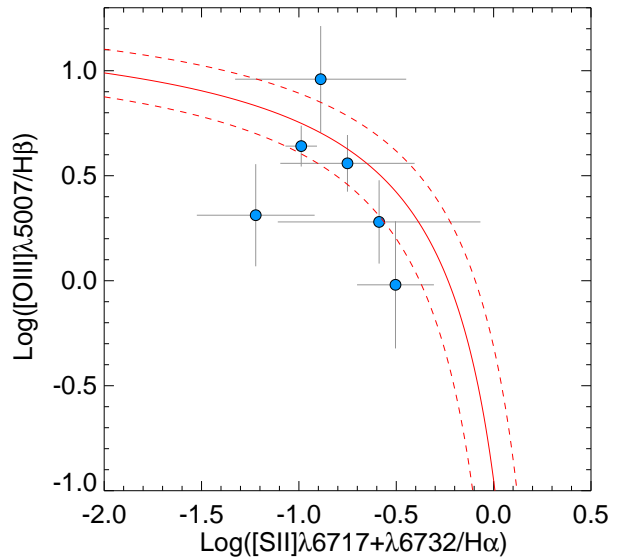} 
   \caption{ The [\oiii] $\lambda5007$/\hb\ versus [\sii] ($\lambda6717 + 6732$)/\ha\ diagnostic diagram. Six of our high-EW galaxies are represented with blue circles with 1$\sigma$ error bars circles. The solid red curve is the separation between star-forming galaxies and AGNi of \citet{kewley01}. The red dashed lines show typical model uncertainties of $\pm$ 0.1 dex}
   \label{fig:bpt}
\end{figure}

The classification of the six objects is presented in Figure \ref{fig:bpt}. We identified one AGN in a subsample of six galaxies that fall in the right redshift range ($0.7 < z < 1.4$) and have detectable \hb\ and [\sii] lines. This represents a fraction of about 17 \%, although the statistics are too small to derive a secure AGN fraction for the entire sample. This is consistent with the fraction found in $z \sim 0.3-2$ \lya\ emitters \citep{atek09b,scarlata09,nilsson09,cowie10, cowie11}, and what is locally found in emission-line galaxy samples \citep[e.g.][]{hao05}.

This high-EW selection is comparable to that of ground-based narrowband surveys \citep{kakazu07}. Also, using broad-band excess, \citet{shim11} identified strong \ha\ emitters in the redshift range of $3.8<z<5.0$ that have mostly EW(\ha)$> 200$ \AA. A total of 176 objects satisfy the high-EW criterion, spanning a redshift range of $0.35 < z < 2.3$, which results in a surface density of 1 object per square arcmin. We are also interested in the redshift evolution of the space density of the high-EW galaxies. In order to compare our results with the local universe we analyzed the SDSS sample of emission-line galaxies at $z < 0.1$. We used the spectral measurements in catalog of the MPA-JHU DR7 release\footnote{\url{http://www.mpa-garching.mpg.de/SDSS/DR7/}} to derive absolute $R$ magnitudes from the SDSS apparent $r$ magnitudes using \citet{jester05} conversion solutions. A passive evolution correction was applied to the WISP sample in order to compare the space densities at a fixed stellar mass \citep[e.g.][]{sargent07}. We applied an absolute magnitude cut at $-21 < M_{R} < -18.5$ in the redshift interval $0.05 < z < 0.1$. The same absolute magnitude cut was applied to the high-EW sample. In the end, the space density of galaxies with EW $> 200$ \AA\ is about $5.5 \times 10^{-6}$ Mpc$^{-3}$ in the SDSS, and about $7 \times 10^{-5}$ Mpc$^{-3}$ in WISP. This represents an evolution of more than a factor of 10 in the number density of strong emission-line galaxies between $z \sim 1.3$ and $z \sim 0$.

\section{Contribution of Emission Lines to Broadband Photometry}
\label{sec:neb_contrib}

\begin{figure}[!htbp]
\centering
   \includegraphics[width=9.5cm]{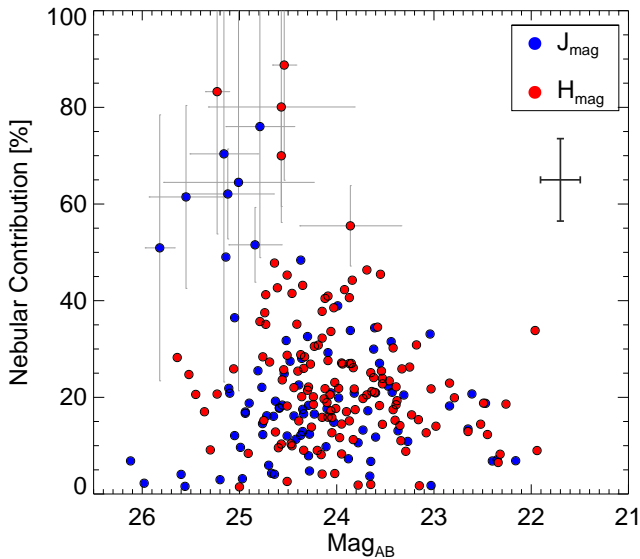} 
   \caption{Contribution of the nebular line flux density to the total broadband flux density in the $J$ and $H$ bands as a function of the total magnitude for objects with EW$_{rest} > 200$ \AA. This quantity is calculated by comparing the synthetic flux density of the emission lines only to the total flux density in the filter. The contribution to the F110W filter flux density is presented in blue and the F140W in red. We plot error bars for objects that have a nebular contribution higher than 50 \%. For object with less than 50 \% contribution, we plot instead an averaged error bars on the top-right of the figure.}
   \label{fig:neb_contrib}
\end{figure}

To quantify the contribution of nebular lines to the total broadband flux density, we have calculated synthetic flux densities from our spectra using the transmission curves of  $J_{110}$ and $H_{140}$ filters and the continuum-subtracted emission lines only, which were then compared to the total flux density. The result is shown in Figure \ref{fig:neb_contrib} where we plot the contribution of the line flux density to the $J_{110}$ and $H_{140}$ total flux densities as a function of the magnitudes in these filters. The nebular contribution can represent most of the broadband flux density of the galaxy, i.e. more than 50 \%, which translates into more than 0.75 mag. The distribution has a median value of about 25 \%, which represents 0.3 mag contribution from the emission lines. The presence of such strong emission lines in the rest-frame optical spectra can have several implications on the study of high-redshift galaxy populations.

\begin{figure*}[htbp]
   \includegraphics[width=9cm]{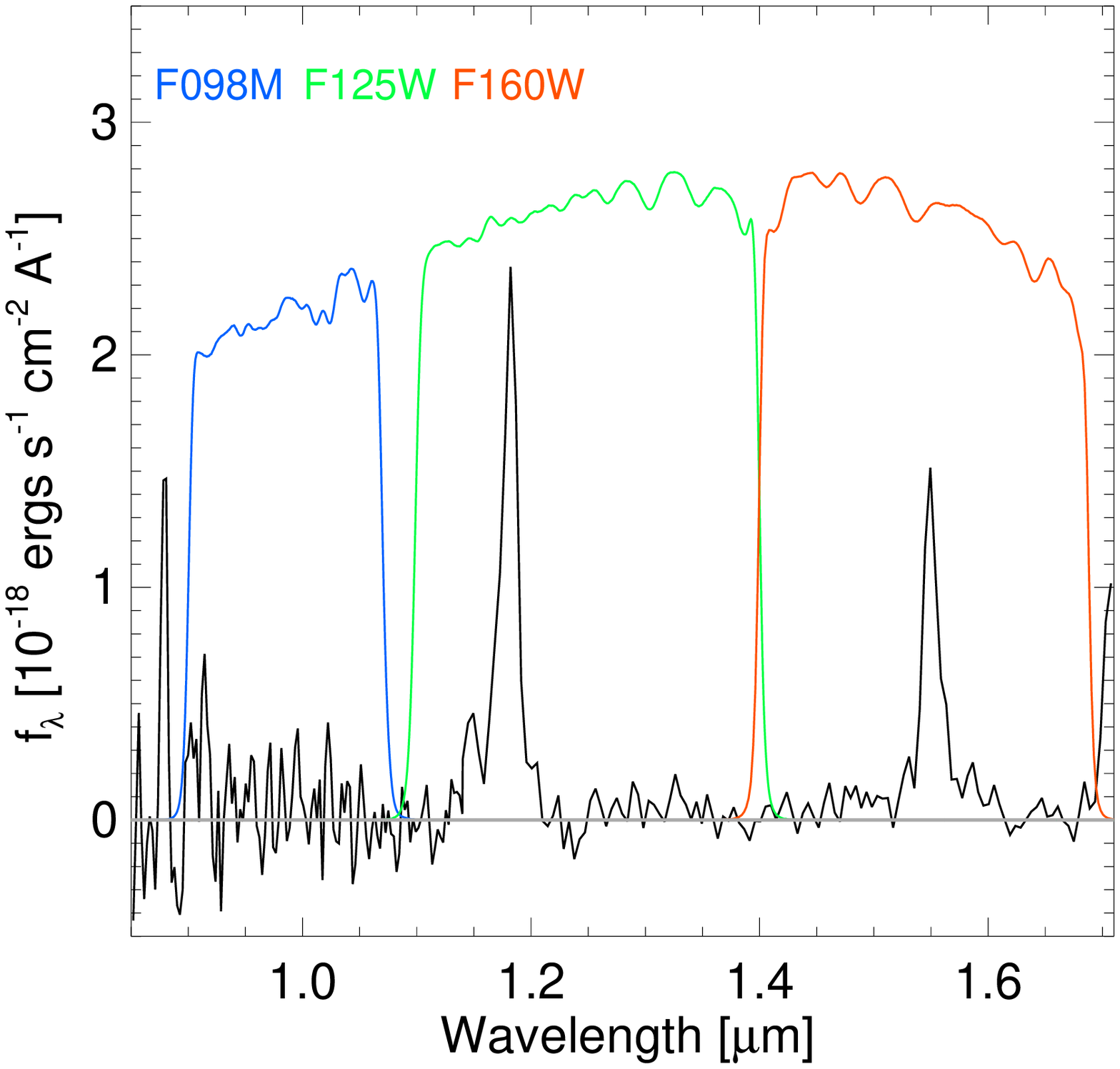}  
   \includegraphics[width=9cm]{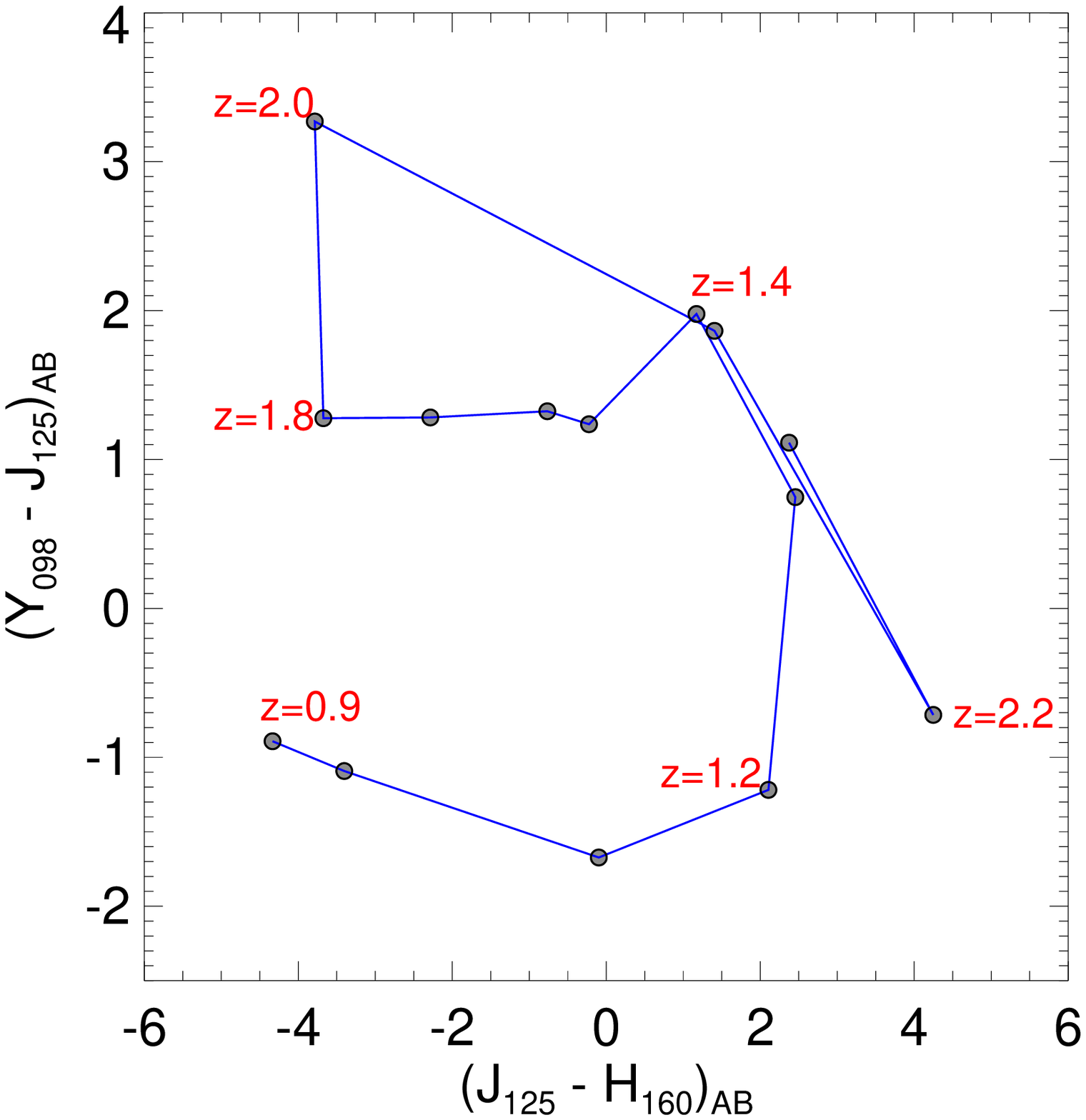}    
   \caption{Contamination of the selection of high-redshift dropout galaxies by lower-redshift sources with strong emission lines. {\it Left:} Example of a high-EW emission-line object, WISP42$\_$115 at redshift $z = 1.36$. The throughput curves of the $098M$, $F125W$, and $F160W$ filters used for the synthetic flux calculations are over plotted. 
   {\it Right:} $Y_{098}-J_{125}$ vs $J_{125}-H_{160}$ color track of strong emission-line galaxy as a function of redshift. An example of a high-EW galaxy spectrum is used to calculate synthetic magnitudes in the IR bands, with EW([\oii] $\lambda$3727, [\oiii] $\lambda$5007, \ha) $\sim$ (200, 800, 550). Beyond the wavelength limits of the WISP spectrum, a flat continuum in $f_{\nu}$ is assumed.} 
   \label{fig:color_track}
\end{figure*}

\subsection{Contamination of high-redshift galaxy samples}
\label{sec:contam}

 In order to select very high-redshift galaxies, the Lyman break technique exploits the continuum drop blueward of \lya\ caused by the intervening IGM absorption. By using a set of broad-band filters, one can sample the continuum of the galaxy to search for a detection in the red band and an absence of signal in the bluer bands. This method has been extensively used to select $z \sim7-10$ candidates using, in particular, the WFC3 IR camera onboard HST \citep{oesch10a,bunker10,yan10,bouwens11,lorenzoni11}. Other groups have identified high-$z$ sources in the WFC3 imaging data set by using SED fitting approach to obtain photometric redshift distributions \citep{mclure10,finkelstein10}. Also, in order to probe the bright end of the luminosity function (LF) at $z \sim 7$, a wide area coverage is necessary. Two {\em HST} observing programs started recently to search for bright $z > 7$ dropout candidates using pure parallel WFC3 NIR imaging of random fields: BoRG \citep[Brightest of Reionizing Galaxies, ][]{trenti11} and HIPPIES \citep[Hubble Infrared Pure Parallel Imaging Extragalactic Survey][]{yan11}.

However, low-redshift sources can have similar colors to those of high-z galaxies and therefore satisfy the dropout selection criteria. The possible contamination from cool dwarfs, transient objects, old galaxies or spurious detections have been addressed in previous studies \citep[e.g.][]{bouwens11}. We consider in this work another potential source of contamination: the strong emission-line galaxies at lower redshift. Since the WISP emission-line galaxy sample is spectroscopically selected over a wavelength range of $0.8-1.7 \mu m$, we were naturally interested in the case of $z \sim > 7$ galaxies, so called $Y$ and $J$ dropouts. We note that the wide shallow observations of \citet{trenti11} and \citet{yan11} use the $Y_{098}$ filter, whereas the deep programs use $Y_{105}$. This can slightly change the results presented hereafter because of the different width of the filters and the wavelength position of the emission lines.

 In order to estimate the contamination of such sources to the dropout selection we use the same filter set used by the wide surveys, i.e. $Y_{098}$, $J_{125}$ and $H_{160}$, and compare their colors. The dominating contributions that affect the broadband colors are [\oiii]$\lambda$$\lambda$5007,4959 and \ha\ emission lines (cf. left panel of Figure \ref{fig:color_track}). In the right panel of Figure \ref{fig:color_track}, we examine the location of the emission lines and their relative contribution to the total flux density of the filters as a function of redshift. Depending upon the wavelength position of the emission lines, a galaxy can appear to have a wide range of IR colors and scatter into the high-$z$ dropout selection window.

 \begin{figure}[!htbp]
 \centering
      \includegraphics[width=9.5cm]{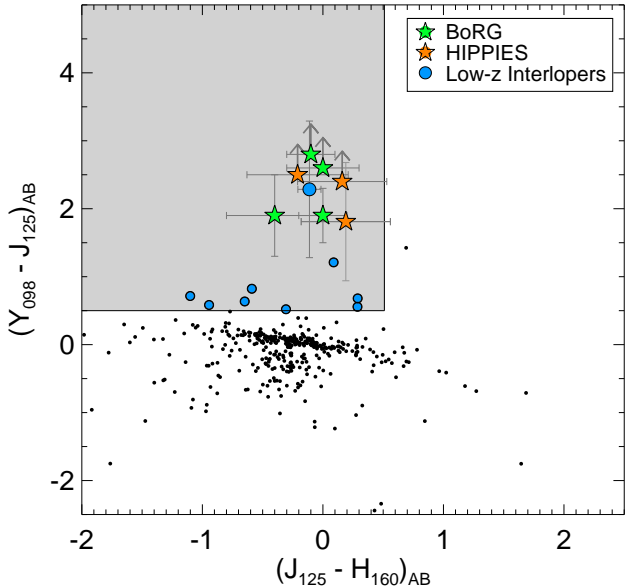}  
   \caption{$Y_{098} - J_{125}$ vs. $J_{125}- H_{160}$ color-color diagram for the high-EW sample of galaxies. The magnitude in each band is derived from synthetic flux calculation using the grism spectra and the WFC3 filter curves. The shaded region denotes the selection window of $z \sim 8\ Y_{098}$-dropout surveys, which satisfy $Y_{098} - J_{125} > 0.5$ and $J_{125}- H_{160} < 0.5$. The galaxies marked in blue circles satisfy this general color-color selection criteria, whereas the rest of sample is shown with black points. In addition, the galaxy represented with a bigger blue circle satisfies $Y_{098} - J_{125} > 1.3$ and $J_{125}- H_{160} < 0.3$ used by the wide surveys. We compare this galaxy with the $Y_{098}$-dropout candidates of \citet{yan11} and \citep{trenti11}, shown in orange and green stars, respectively, and whose color criteria are detailed in the text. The error bars indicate 1$\sigma$ uncertainties, while the arrows represent lower limits.}
   \label{fig:ydrop}
\end{figure}

 \begin{figure*}[!htbp]
  \hspace{-4cm}
   \includegraphics[width=17cm]{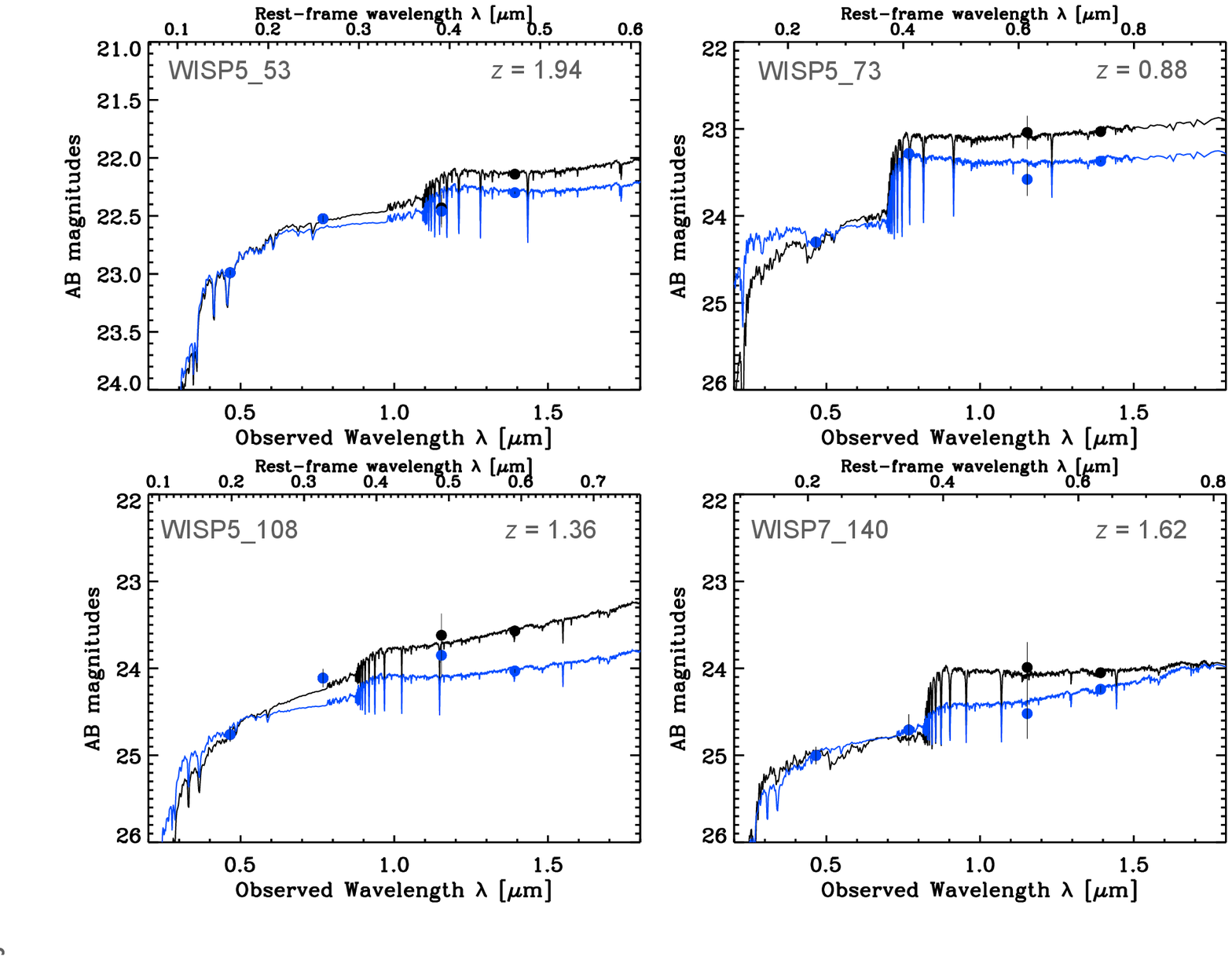}  
   \vspace{-5cm}
   \caption{SED fitting of the high-EW galaxies. The black points represent the observed magnitudes in the $g', i', J,$ and $H$ bands, while the blue points are the magnitudes corrected for the contamination of nebular lines following the procedure explained in the text. The error bars show 1$\sigma$ uncertainties. The black and blue curves are the best fit models to the raw and corrected magnitudes, respectively. We used \citet{bc03} models assuming exponentially declining star formation history, 0.02 $Z_\odot$ metallicity, and \citet{chabrier03} IMF.} 
   \label{fig:sed1}
\end{figure*}

\begin{figure*}[!htbp]
  \hspace{-4cm}
   \includegraphics[width=17cm]{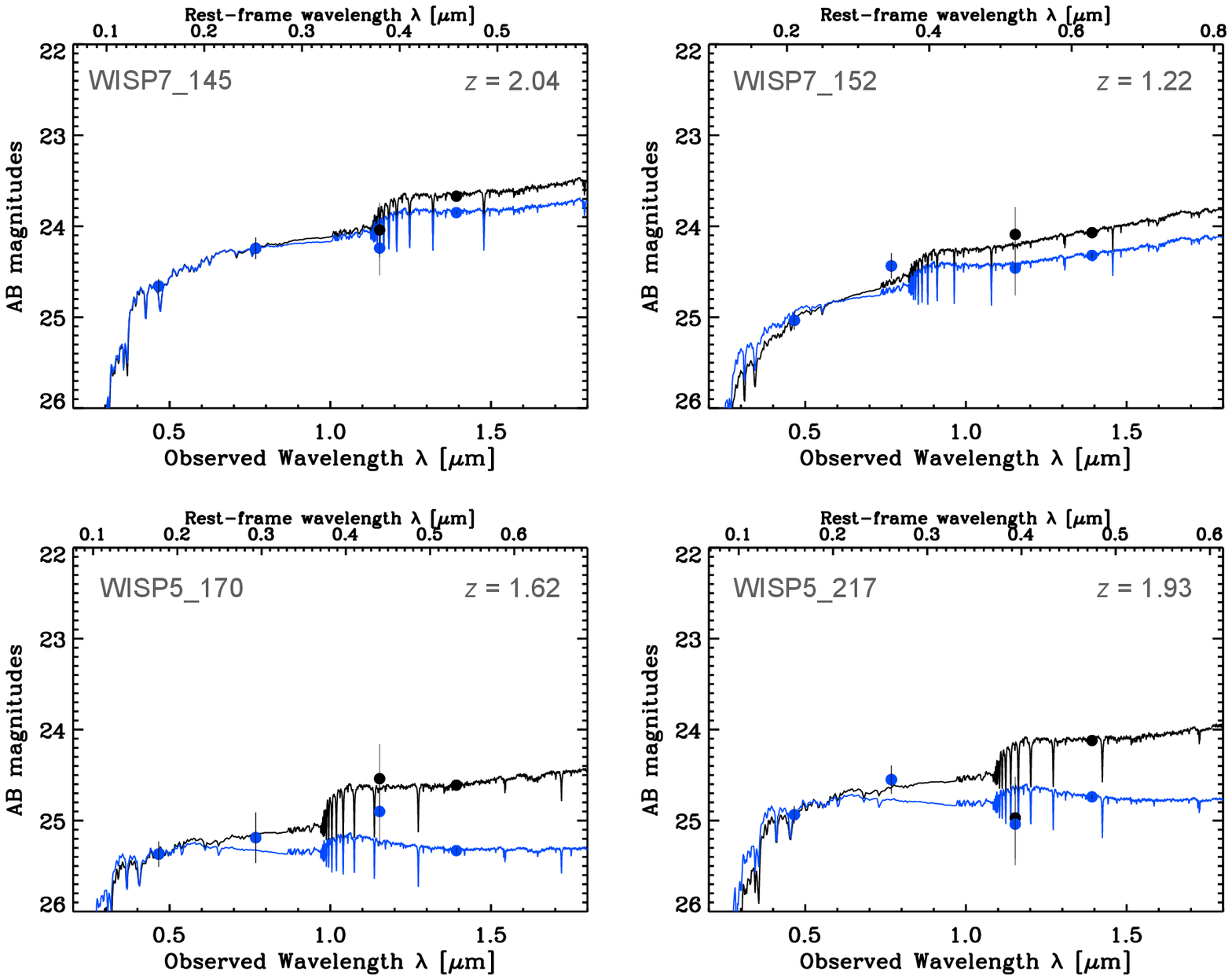}  
   \vspace{-5cm}
   \caption{Same as Figure \ref{fig:sed1}.} 
   \label{fig:sed2}
\end{figure*}

 Figure \ref{fig:ydrop} shows the $Y_{098} - J_{125}$ vs. $J_{125}- H_{160}$ color-color diagram commonly used to isolate high-$z$ candidates populated with the high-EW galaxies. We find a large dispersion in both $J_{125}- H_{160}$ and $Y_{098} - J_{125}$ colors, and we observe galaxies that fall in the high-z selection window (grey region) that denotes the color selection $J_{125}- H_{160} < 0.5$ and $Y_{098} - J_{125} > 0.5$ \citep[e.g.][]{bouwens10}.

The contamination of $z \sim 8$ galaxy candidates by low-$z$ interlopers with strong emission lines has been investigated theoretically by \citet{taniguchi10}. These authors used SED models to predict the strength of the emission lines from the ionizing continuum, and then examined whether their synthetic spectra satisfy the selection criterion of $z \sim 8$ galaxies. The [\oiii] $\lambda$5007 equivalent width in their models varies from $\sim 10^{3}$ \AA\ to $\sim 100$ \AA\ for age of 1 Myr to 100 Myr. They conclude that the contribution of the emission lines is too small to meet the color requirement. We measure $EW > 1000$ \AA\ in $z \sim 1-2$ galaxies, however, and these do satisfy the Y-dropout selection as we can see in Figure \ref{fig:ydrop}. However, one of the main conditions used to rule-out low$-z$ interlopers in such surveys is the non-detection in the optical domain. To this end, HUDF observations \citep{bouwens10,bunker10,yan10} make use of ACS data in several optical bands which are at least 1 magnitude deeper than IR images. This makes the contamination from low-$z$ strong line emitters very unlikely, unless the extinction is very high. We were able to measure the reddening from \ha/\hb\ ratio for one of the interlopers of Figure \ref{fig:ydrop} (purple circles) and found a small value of E(B-V) $\sim$ 0.05 (A$_{v} \sim 0.15$). Also, in the rest of the high-EW sample, objects for which \ha\ and \hb\ are available show little extinction. It is also possible that this is the result of a selection effect in the sense that we are missing the dusty galaxies where \hb\ line remains undetectable because it would be fainter than our flux limit.

On the other hand, the wide pure parallel surveys BoRG and HIPPIES do not use such deep optical data to discard these interlopers. Indeed, the UVIS observations in the $F606W$ or $F600LP$ bands reach a depth comparable to the IR. The color selection criteria used in these studies are $Y_{098} - J_{125} > 1.75$ and $J_{125}- H_{160} < 0.02 + 0.15 \times (Y_{098} - J_{125} - 1.75)$ for \citet{trenti11}, and $Y_{098} - J_{125} > 1.35$  and $J_{125}- H_{160} < 0.3$ for \citet{yan11}, respectively. We plot in Figure \ref{fig:ydrop} the colors of the high-EW galaxies (in blue) together with the samples of $z \sim 8$ candidates selected in the two parallel surveys (orange and green).  When applying the respective color criteria of these two surveys, and given the uncertainties, we end up with 1 interloper in both the BoRG and HIPPIES samples. In this section, we used 24 fields that have both \grisma\ and \grismb\ observations, which yields an effective survey area of  about 80 acrmin$^{2}$. When we scale this area to the above surveys, we find that the contamination from such sources represents about 1 object in every 17 fields observed by the wide shallow surveys. This is assuming the same number density of such sources at the depth of the wide surveys, because their photometric observations go deeper than our spectroscopic survey.

To show the necessity of deep optical imaging, we have estimated the expected flux in the optical bands of these interlopers using the best fit SEDs of the high-EW galaxies presented in Sect. \ref{sec:sed}. Using the filter throughputs, we calculated the magnitudes in the $F606W$ and $F600LP$ optical bands. We find that almost all the galaxies have red $V_{606}-Y_{098}$ colors ranging from 0.1 to 0.9 mag. The same red colors are obtained when we use the $F600LP$ filter. These results are also confirmed by the $g' - J$ and $i' - J$ colors observed for these objects (see Table \ref{tab:mag}). While some of the interlopers that satisfy the IR color criteria can be identified in the UVIS filters, a mean value of $V_{606}-Y_{098} \sim 0.6$ in our results suggests that $\sim 1$ mag deeper imaging is needed in the optical in order to rule out the contamination from all the low-$z$ high-EW galaxies. In this sense, the ongoing wide survey of the Cosmic Assembly Near-IR Deep Extragalactic Legacy Survey \citep[CANDELS,][]{grogin11} will have appropriately deep optical imaging in multiple {\em HST}/ACS (Advanced Camera for Surveys) bands.

\begin{deluxetable*}{rcclcccc}
\centering
\tablecaption{Broadband Photometry of the Sample \label{tab:mag}}
\tablehead{
\colhead{Object} & $z$  &\colhead{$g'$} & \colhead{$i'$} & \colhead{$J$} & \colhead{$J_{corr}$} & \colhead{$H$} & \colhead{$H_{corr}$}  \\
\colhead{$ $}      &    &\colhead{(mag)} & \colhead{(mag)} & \colhead{(mag)} & \colhead{(mag)}& \colhead{(mag)} & \colhead{(mag)} \\
}
\startdata
WISP5$\_$53    &  1.94  & 22.98 (0.02)  & 22.52 (0.04)  &  22.43 (0.14)  & 22.46  & 22.14 (0.01)  & 22.30 \\
WISP5$\_$73    &  0.88  & 24.30 (0.06)  & 23.28 (0.05)  &  23.04 (0.19)  & 23.58  & 23.03 (0.01)  & 23.37 \\
WISP5$\_$108  &  1.36  & 24.76 (0.08)  & 24.11 (0.10)  &  23.62 (0.25)  & 23.85  & 23.57 (0.02)  & 24.03 \\
WISP5$\_$170  &  1.62  & 25.36 (0.14)  & 25.19 (0.28)  &  24.54 (0.38)  & 24.90  & 24.61 (0.04)  & 25.33 \\
WISP5$\_$217  &  1.93  & 24.93 (0.09)  & 24.55 (0.16)  &  24.97 (0.45)  & 25.04  & 24.12 (0.02)  & 24.74 \\
WISP5$\_$230  &  0.70  & 25.93 (0.23)  & 25.84\tablenotemark{a} (0.16)  &  25.12 (0.48)  & 26.17  & 25.52 (0.08)  & 25.53 \\
WISP7$\_$140  &  1.20  & 25.00 (0.10)  & 24.70 (0.18)  &  23.99 (0.29)  & 24.52  & 24.05 (0.04)  & 24.24 \\
WISP7$\_$145  &  2.04  & 24.65 (0.07)  & 24.24 (0.12)  &  24.04 (0.30)  & 24.24  & 23.67 (0.02)  & 23.85 \\
WISP7$\_$152  &  1.22  & 25.03 (0.10)  & 24.43 (0.14)  &  24.09 (0.30)  & 24.46  & 24.07 (0.03)  & 24.32 \\
\enddata
\tablecomments{The optical and near-infrared magnitudes of our subsample of strong emission-line galaxies. The $J$ and $H$ magnitudes are from our WFC3 IR observations, while the $g'$ and $i'$ data were obtained during our optical follow-up at Palomar observatory. The $J_{corr}$ and $H_{corr}$ columns correspond to the magnitude values corrected for emission-line contribution. The 1-$\sigma$ uncertainties are given in parentheses. The ground-based photometry aperture was corrected to match that of the WFC3 images (cf. Section \ref{sec:follow}). Magnitudes are in AB system.}
\tablenotetext{a}{The $i'$ band flux of this object was corrected (by 1.3 mag) for the contamination of strong emission lines using the LRIS spectrum.}
\end{deluxetable*}

\begin{deluxetable*}{r c c c c c c}
\centering
\tablecaption{Impact of emission lines on the physical properties \label{tab:sed}}
\tablehead{
\colhead{Object} &  \colhead{$A_{v}$}&  \colhead{log(SFR)} &  \colhead{log(Age)} & \colhead{log(Age (corr)) } &  \colhead{log(Mass)}  & \colhead{Mass (corr)}  \\
\colhead{ }          &                                   & \colhead{(\msolyr)}  & \colhead{(yr)}         &  \colhead{(yr)}                  & \colhead{(\msol)}             & \colhead{(\msol)}  \\
}
\startdata
WISP5$\_$53   &1.20 & 2.58    & 7.00 & 7.00    &9.64 & 9.52   \\
WISP5$\_$73   &0.20 &-0.90    & 8.20 & 8.40    &9.08 & 8.91   \\
WISP5$\_$108  &1.60 & 1.83    & 7.00 & 7.00    &8.89 & 8.56   \\
WISP5$\_$170  &0.70 & 0.78    & 8.20 & 7.00    &8.90 & 8.00   \\
WISP5$\_$217  &0.80 & 1.23    & 8.00 & 7.00    &9.21 & 8.35   \\
WISP5$\_$230  &0.00 &-1.00    & 8.30 & 8.20    &7.81 & 7.67   \\
WISP7$\_$140  &0.20 &-0.54    & 8.10 & 7.20    &8.89 & 8.43   \\
WISP7$\_$145  &1.00 & 1.63    & 7.90 & 7.30    &9.45 & 9.09   \\
WISP7$\_$152  &1.50 & 1.57    & 7.00 & 7.00    &8.55 & 8.34   \\
\enddata
\tablecomments{Physical properties derived from the SED fitting. The best fit for the age and and the stellar mass are presented with (corr) and without including the nebular emission lines.}
\end{deluxetable*}

The high-z narrow band surveys of \lya\ emitters can also be, to some extent, subject to the contamination of these strong low-$z$ emitters. The selection method relies on the flux excess detected in the narrowband relative to the broadband flux, and selects candidates that have rest-frame equivalent width higher than $\sim$ 40 \AA. Obviously, the galaxies presented in this work have much larger EW values, and the narrowband excess can be attributed in some cases to these lower redshift interlopers. However, at redshift $z \sim 6.5$ for instance, \citet{ouchi10} and \citet{kashikawa11} use also the color blueward of the \lya\ emission to detect the Lyman break of the NB921-selected galaxies. This additional criterion significantly reduces the low-$z$ contamination fraction. Using $ R-i > 1$ and $i-z > 1.2$, \citet{ouchi08,ouchi10} found in their spectroscopic follow-up that such contamination is very small for their sample of $z \sim 5.7$ and $z \sim 6.6$ candidates, respectively. Using similar constraints, \citet{kashikawa11} report 81\% (70\%) confirmation rate for their $z=6.5$ (5.7) LAE candidates. Combining the color check blueward of \lya\ with the narrowband selection should be sufficient to identify the low-$z$ high-EW sources.

\subsection{Effects on Age and Mass Estimate of High-z Galaxies}
\label{sec:sed}

While most stellar population synthesis models do not include the nebular emission lines, some effort has been devoted recently to take into account the impact of such contribution on the SED analysis of high-z galaxies. \citet{eyles07} used the UV star formation rate to assess the contribution of optical emission lines to their $Spitzer$ IRAC photometry and their effect on the SED fitting of $z \sim 6$ galaxies. More recently, \citet{finkelstein11} corrected for the optical emission lines in their fitting of two LAEs at $z \sim 2.3$ \citep[see also][]{mclinden11} . \citet{schaerer09} treated the effect of nebular emission (lines and continua) in a self-consistent way by predicting the absolute line intensities from the ionizing continuum of the template SED. They showed that neglecting the nebular component could lead to overestimate the age of $z \sim 6$ galaxies by a factor of 3. This theoretical prescription has been followed in several papers \citep[eg.][]{ono10,watson10,raiter10,finlator11,mclure11,inoue11} dealing with SED modeling.

Using the sample of very-high EW objects, we empirically study the impact of nebular emission lines on the SED fitting of galaxies in general, and on the age and mass estimates in particular. We selected 9 galaxies from fields WISP5 and WISP7 that have optical follow up in $g$ and $i$ bands obtained with the 5m {\em Hale} telescope at Palomar observatory, in addition to the $J$ and $H$ photometry from the WISPS IR observations.

To model the stellar population properties of these galaxies, we use the {\tt FAST} code \citep{kriek09} to fit \citet{bc03} synthesis models to the continuum magnitudes. We assume exponentially declining star formation history, a metallicity of $Z = 0.02$, and a \citet{chabrier03} initial mass function (IMF). The fitting results are presented in Figure \ref{fig:sed_230}, \ref{fig:sed1}, and \ref{fig:sed2}. The black curve is the model spectrum fit to the raw observed magnitudes. The blue curve is the model fit to the magnitudes corrected for the contamination of nebular emission lines following the procedure detailed in Section \ref{sec:contam}. In addition, we used the LRIS spectrum to correct the $i'$ band magnitude of WISP5$\_$230. Depending on the wavelength location and the strength of the contributing emission lines, the difference between the models varies dramatically. 

In Table \ref{tab:sed}, we compare the stellar populations properties derived from the two model fits. It is shown that not accounting for the contribution of emission lines can significantly affect the age and the stellar mass estimates in these galaxies. Specifically, the age and the mass of individual objects can be overestimated by an order of magnitude. We show for instance that, based on the observed magnitudes, the galaxy WISP5$\_$170 has an age of $t \sim 160$ Myr and a mass of $M^{\star} \sim 8 \times 10^{8}$ \msol. Once corrected for the nebular emission, the age and mass are revised to $t \sim 10$ Myr and $M^{\star} \sim 10^{8}$ \msol, respectively. Considering all the objects with varying emission-line intensities, we derive a mean age for the sample of $t_{mean} \sim 100$ Myr and a mass of $M^{\star}_{mean} \sim 1.4 \times 10^{9}$ \msol. Correcting for emission lines brings these values down to $t_{mean} \sim 55$ Myr and a mass of $M^{\star}_{mean} \sim 7.3 \times 10^{8}$ \msol. This change represents an average correction factor of 2. For comparison, \citet{finkelstein11} derived for their two objects at $z \sim 2.5$ a correction factor of (2, 6) for the mass and a factor of (10, 1) for the age. We note that the extinction $A_{v}$ can also be slightly overestimated if no correction is applied.

It is clear that not accounting for nebular emission lines when modeling their SEDs can introduce large errors in the derived age and stellar masses of star-forming galaxies at all redshifts. Deriving the physical properties of higher redshift galaxies is much more problematic. Galaxies at $z > 6$ will be younger, more metal-deficient, and will exhibit higher equivalent widths. Moreover, the contribution of emission lines to the broadband filters increases as ($1+z$). However, this effect is mitigated by the fact that commonly used broad photometric bands increase in width when going redder. This will dilute the line in this large wavelength band and decrease the EW contribution. This empirical demonstration shows the need for careful SED modeling by systematically accounting for nebular emission lines.

\begin{figure}[!htbp]
   \centering
   \includegraphics[width=10cm]{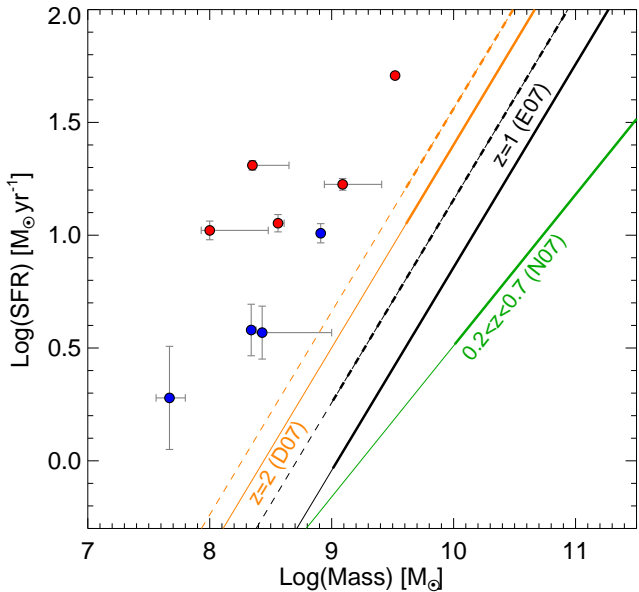} 
   \caption{The star formation rate as a function of the stellar mass. The SFR is derived from the optical emission lines and the stellar masses from the SED fitting of the optical and NIR photometry (cf. text for details). Objects located at $z > 1.3$ are shown with red circles and objects at $z < 1.3$ with blue circles. The 1$\sigma$ uncertainties on the mass and SFR are also overplotted. The lines denote previous results of several studies of the SFR-mass correlation: the orange line is the fit to $z=2$ sample of \citet{daddi07} with the dashed line representing the typical dispersion of 0.16 dex. The black line is for galaxies at $z=1$ from \citet{elbaz07} with a dispersion of 0.3 dex (dashed black line) and the green line for galaxies at $0.2 < z < 0.7$ from \citet{noeske07}. The thick part of each line represents the correlation in the mass range explored by each study and the thin line is an extrapolation to lower masses.}
   \label{fig:sfr_mass}
\end{figure}

\section{High Specific Star formation rate galaxies}
\label{sec:ssfr}

The correlation between the SFR of galaxies and their stellar mass $M^{\star}$ or their luminosity and metallicity has been studied extensively, especially at low redshift. Observationally, it has been investigated at different redshifts \citep{brinchmann04,erb06,noeske07, elbaz07, daddi07,damen09,labbe10}. This SFR$-M^{\star}$ relationship is sometimes called the galaxy main sequence \citep{noeske07} as a reference to the stellar main sequence. While a SFR$-M^{\star}$ relation is exhibited at all redshifts, it evolves in the sense that galaxies of a given stellar mass have higher SFRs at higher redshifts. The evolution of the SFR as a function of $M^{\star}$ is of particular interest as a test of galaxy evolution models. The SFR-$M^{\star}$ correlation is indeed predicted by cosmological hydrodynamic simulations \citep{finlator11}, and \citet{bouche10} argue that smooth gas accretion is responsible for this trend \citep[see also][]{dave11}

However, at high-redshift, the SFR$-M^{\star}$ correlation has been limited to a relatively high mass range. The techniques used for the selection of galaxies include mostly the optical or NIR colors and UV continuum \citep[e.g.][]{steidel04,daddi04,papovich06,steidel96,shapley05,erb06,schiminovich07}. Unfortunately, this misses a population of star-forming galaxies with faint continua. Photometric redshifts have also been used to identify high-$z$ galaxies, but remain unreliable at faint magnitudes because of photometric uncertainties and emission line contribution. While the most massive galaxies are offset to lower specific star formation rates (sSFR, the SFR per unit stellar mass) in the SFR$-M^{\star}$ plane, 
our sample selection targets precisely galaxies with higher sSFRs that would lie in the upper left corner of this plane.

In Figure \ref{fig:sfr_mass}, we present the star formation rate measured from emission lines as a function of the stellar mass for 9 of the high-EW sources for which we obtained optical follow-up imaging. The SFR is derived from the \ha\ flux using \citet{kennicutt98} calibration. No dust correction was applied, as we were able to measure \ha\ and \hb\ lines in 5 of these galaxies that were consistent with no, or little, extinction. We corrected the \ha\ flux for [\nii] contamination using the procedure described in Section \ref{sec:high_ew}. For high-redshift galaxies selected by their strong [\oiii] $\lambda$5007 line, and where the \ha\ line is outside the \grismb\ window, we also have a detectable [\oii] $\lambda$3727 line. Likewise, we apply the \citet{kennicutt98} prescription to convert the [\oii] flux into SFR. The stellar masses were derived from SED fitting of the optical and NIR observations of the galaxies using \citet{bc03} spectral synthesis models with exponentially declining star formation $e^{-t/\tau}$ with $t$ and $\tau$ as free parameters, a \citet{chabrier03} initial mass function, and metallicity of $Z = 0.02$. 

It can be seen from Figure \ref{fig:sfr_mass} that these galaxies have a higher sSFR than the sequence of normal star-forming galaxies at the same redshift derived in previous studies. They probe a mass range that has never been explored before.  We divided the high-EW sample into two bins of redshifts below and above $z = 1.3$ drawn in blue and red, respectively. While \citet{daddi07} derive a tight SFR-$M^{\star}$ correlation for $z \sim 2$ galaxies in GOODS (Great Observatories Origins Deep Survey), with a dispersion of 0.16 dex in SFR, the present galaxies lie well above the 0.5 dex line. They represent extreme outliers to the relationship at $z \sim 2$, like the sub-millimeter galaxies (SMGs) but orders of magnitude smaller and less massive, because we probe stellar masses in the range $10^{7.5} < M^{\star} < 10^{9.5}$ \msolyr, much lower than \citet{daddi07}. Compared to the $z \sim 1$ GOODS sample \citep{elbaz07}, for which the SFR-$M^{\star}$ relationship has 0.3 dex dispersion, the high-EW sample of galaxies are offset by more than 1 dex from the median value of SFR at a given mass. \citet{noeske07b} interpret $z \sim 1$ galaxies on the SFR-$M^{\star}$ main sequence as the early phase of a star formation history (SFH) that smoothly declines for $\sim 1$ Gyr to $z \sim 0$.ÊHowever our observed sSFRs denote vigorous star formation episodes with very rapid stellar mass build-up, doubling total stellar mass in $< 100$ Myr. The observed offset of our galaxies from the SFR-$M^{\star}$ main sequence is instead consistent with strong burst episodes rather than a smoothly declining SFH.

\section{Extremely Low Metallicity Galaxies}
\label{sec:low_metal}

\begin{figure*}[!htbp]
   \centering
     \vspace{-3.6cm}
   \includegraphics[width=13cm]{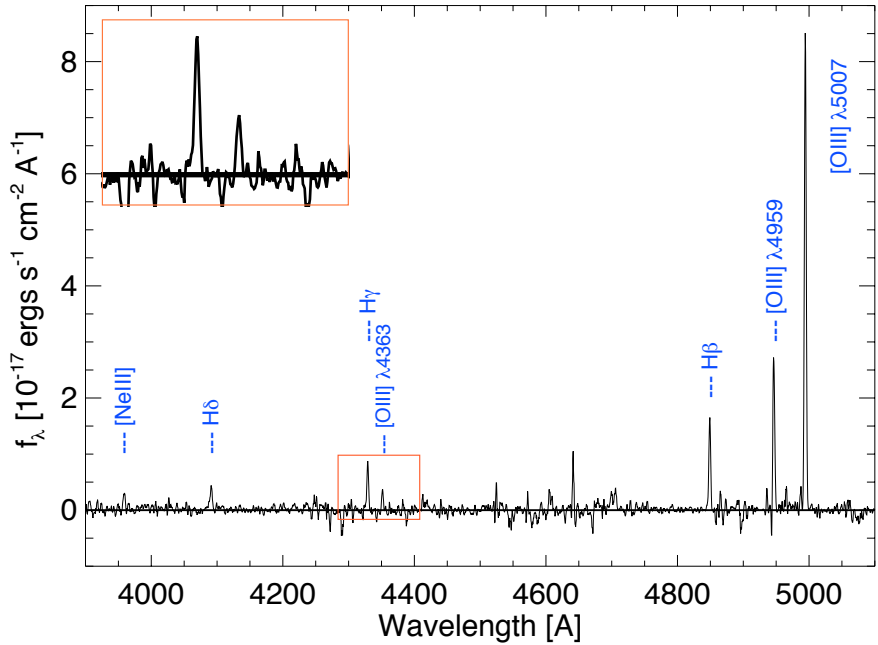} 
   \vspace{-3.6cm}
   \caption{LRIS rest-frame spectrum of the XMPG WISP5$\_$230 at $z \sim 0.7$. We observe a very faint continuum with very strong emission lines, such as \hb\ or [\oiii]$\lambda$5007. The inset shows a close-up of the H$\gamma$ and [\oiii] $\lambda$4363 lines. The detection of the [\oiii] $\lambda$4363 auroral line suggests a very low metallicity. The spectrum has been smoothed by 3-pix filter.}
   \label{fig:239}
\end{figure*}

It has long been known that lower luminosity galaxies tend also to have lower metallicities according to the luminosity-metallicity relationship, although with a large scatter \citep{tremonti04,salzer05,savaglio05,erb06}. Extremely metal-poor galaxies (XMPGs) with oxygen abundance of log(O/H) + 12 $ \le 7.65$ as defined by \citet{kunth00} are very rare \citep[e.g.][]{morales-luis11}. In the local universe, large EWs galaxies have been surveyed using broad-band color selection \citep{brown08}. \citet{cardamone09} found compact star-forming galaxies in the SDSS, called ``green peas'', that have very blue $r - i$ colors because of their very strong [\oiii]$\lambda$5007 emission line. \citet{kakazu07} showed that emission line selection is yet more efficient than broad-band selection and subsequent followup observations \citep[e.g. DEEP2][]{davis03, hoyos05}.

Many of the faint WISP galaxies have strong [\oiii] $\lambda$5007 emission, and often weak [\oii] $\lambda$3727 emission, with a median ratio of  ([\oiii] $\lambda$5007/[\oii] $\lambda$3727) = 2.5 \citep[see also][]{hicks02, maier06, ly07}, suggestive of low metal abundances (Log(O/H) $\lesssim 8.5$)). We present here an example of a high-EW galaxy that shows a very low metallicity. 

Amongst 5 high-EW galaxies for which we obtained LRIS spectra, only one galaxy was at a suitable redshift to observe the optical emission lines necessary for the metallicity measurement. The galaxy WISP5$\_$230 proves to be an XMPG. It lies at a redshift of $z = 0.7$ and its LRIS spectrum (Figure \ref{fig:239}) confirms very strong emission lines seen in the grism spectra, with no detected continuum. In particular, the [\oiii] $\lambda$5007 line has an equivalent width lower limit of EW$_{rest}  >  860$ \AA, and the \hb\ line has EW$_{rest} > 180$ \AA. We also detect the [\oiii]$\lambda$4363 auroral line, the presence of which always indicates a low metallicity, with $f$([\oiii]$\lambda$4363) $= 4.0 \pm 0.3 \times 10^{-18}$ \ergscm, which is almost half of the flux of the H$\gamma$ line. We used the direct method to derive the metallicity using the ratio of [\oiii] $\lambda$5007,4959 and [\oiii]$\lambda$4363 that allows us to measure the electron temperature $t$.  Using the {\tt NEBULAR} package in {\tt IRAF} \citep{shaw95}, and assuming an electron density $n_{e}$ of 100 cm$^{-3}$, we compute an electron temperature $t \sim 2 \times 10^{4}$ K. We finally derive the oxygen abundance following the equations of \citet{izotov06}:

\begin{eqnarray}
12+\log {\rm O}^{+}/{\rm H}^+=&\log \frac{\lambda 3727}{{\rm 
H}\beta}+5.961+\frac{1.676}{t}
-0.40\log t \nonumber\\
                  &-0.034t+\log (1+1.35x),
\label{oii1}
\end{eqnarray}
\begin{eqnarray}
12+\log {\rm O}^{2+}/{\rm H}^+=&\log \frac{\lambda 4959+\lambda 
5007}{{\rm H}\beta}+6.200+\frac{1.251}{t}
  \nonumber\\
                  &-0.55\log t-0.014t
\label{oiii}
\end{eqnarray}
where  $x = 10^{-4} n_{e} t^{-0.5}$. Since the [\oii]$\lambda$3727 line fell on a sky line, we first use a fit to local \hii\ galaxies used in \citet{hoyos05}, which gives the ([\oiii] $\lambda$5007/[\oii] $\lambda$3727) ratio as a function of \hb\ equivalent width, where the typical uncertainty on the [\oii] flux is about 50\%. Then, using the typical ratio of 2.5 observed for the WISP galaxies we obtain a [\oii] flux higher by 50\% than the previous estimate. We eventually adopt the typical [\oiii]/[\oii] line ratio observed in WISP as the final value. In fact, the uncertainties do not strongly depend on the precise value of the [\oii] flux. The flux errors were propagated during the calculation. Finally, we obtain an oxygen abundance of 12+log(O/H)=$7.47 \pm 0.11$. For comparison, the solar metallicity of $Z=0.02$ translates to 12+log(O/H) $\sim 8.9$.

\begin{figure}[!htbp]
   \centering
   \includegraphics[width=9.8cm]{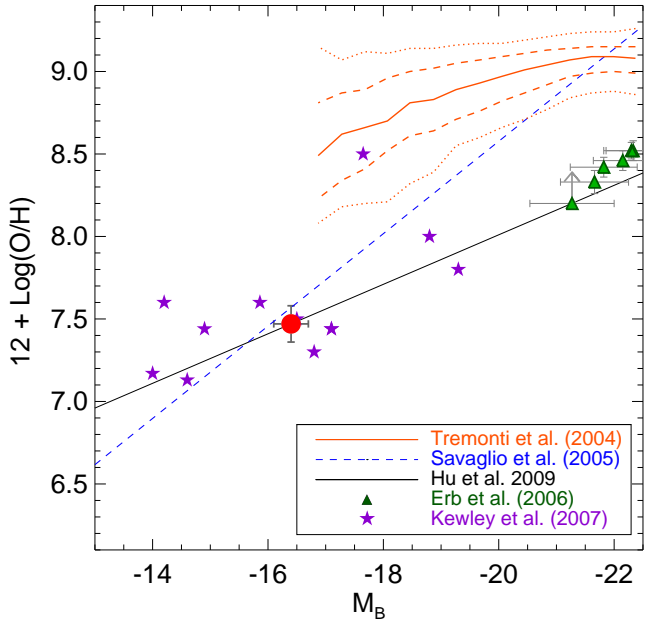} 
   \caption{The high-EW emission-line selection is uncovering extremely low metallicity objects. We show here the position of WISP5$\_$230 in the metallicity-luminosity plane, together with results from the literature. The solid orange line represents the median of the low-redshift sample in SDSS of \citet{tremonti04}, while the dashed and dotted orange lines are the contours that enclose 68\% and 95\% of the sample. The blue dashed line denotes the $z \sim 0.7$ galaxy sample of \citet{savaglio05} and the green triangles the $z \sim 2.3$ star-forming galaxies of \citet{erb06} divided into six bins of luminosity. Local XMPGs and GRB hosts from \citet{kewley07} are shown in magenta stars and XMPGs of the USEL sample is represented by the black solid line \citep{hu09}. The position of WISP5$\_$230 is marked with a big red circle.}
   \label{fig:metal}
\end{figure}

This object is comparable to the most metal-deficient galaxies found in the USEL sample \citep{hu09} and close to the lowest metallicity star-forming galaxies known, \izw\ \citep{thuan05} and SBS0335-52 \citep{izotov06}. We plot the metallicity and the absolute B-band magnitude of WISP5$\_$230 in Figure \ref{fig:metal} and compare it with results at different redshifts. The luminosity-metallicity relation is an important indicator of the chemical enrichment of the inter-stellar medium and the mass-loss of galaxies. In the same figure, we show the relation established for a large sample of low-redshift star-forming galaxies in the SDSS \citep[orange lines,][]{tremonti04}, at redshifts $0.4<z<1$ \citep[blue dashed line,][]{savaglio05}, and for high-redshift LBGs at $z \sim 2.3$ \citep[green triangles,][]{erb06}. With an absolute magnitude of $M_{B} \sim -16.4$, WISP5$\_$230 is comparable to local XMPGs (magenta stars), including the blue compact galaxy SDSS 0809+1729 found by \citet{kewley07}. 

We observe that, at constant luminosity, it has a metallicity $\sim 1$ dex lower than what would be predicted from local SDSS galaxies \citep{tremonti04}. It is better reproduced by the black solid line in Figure \ref{fig:metal} that represents the same trend for the USEL sample \citep{hu09}. It is also consistent with relationship of \citet{savaglio05} derived for $z \sim 0.7$ galaxies.

\begin{figure}[!htbp]
   \centering
   \vspace{0.3cm}
   \includegraphics[width=8.3cm]{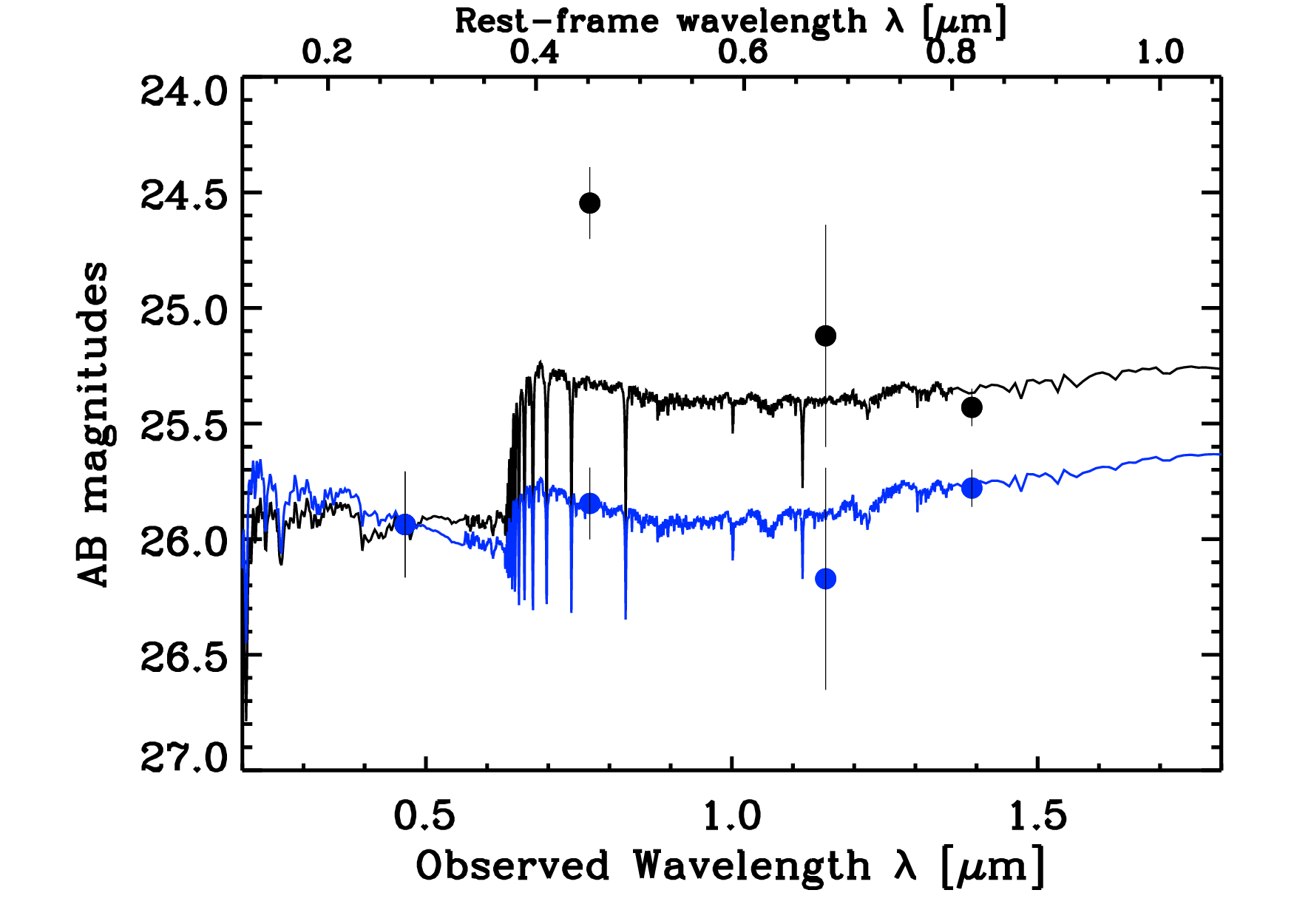} 
   \caption{Spectral energy distribution fitting of XMPG WISP5$\_230$. The black points represent the observed magnitudes in the $g', i', J$ and $H$ bands. The blue points are the magnitudes corrected for the contribution of nebular emission lines. The $J$ and $H$ fluxes are corrected using the WFC3 grism spectra, and the $g'$ flux using the LRIS spectrum. The solid black (blue) curves is the best fit SED to the raw (emission-line corrected) magnitudes using \citet{bc03} models. The SED fitting procedure is detailed in Section \ref{sec:sed}.}
   \label{fig:sed_230}
\end{figure}

This object also has a stellar mass of Log(M$^{\star}$) $= 7.67 \pm 0.13$ \msol; the lowest value in the sample of 9 galaxies for which we performed SED modeling (see Section \ref{sec:sed}). The best-fit SED is shown in Figure \ref{fig:sed_230}. The low metallicity and low mass of this object are consistent with the well known mass-metallicity relation \citep{tremonti04, savaglio05, erb06, mannucci09}. In addition, \citet{mannucci10} argue that  the mass-metallicity relation is simply a projection of a more fundamental relation between mass, gas-phase metallicity and SFR. The evolution of the mass-metallicity relation up to $z 
= 2.5$ would be the result of an evolution within this fundamental relation, as higher-SFR galaxies are being selected at increasing redshifts. Therefore, for low-mass galaxies, the metallicity decreases sharply with increasing SFR. However, the high-redshift samples are still small and there is not enough overlap between SFRs of the low- and high-$z$ samples to confirm this scenario. Using the fundamental metallicity equation derived in \citet{mannucci10} we find that, with SFR$\sim 1.9$ \msolyr, the galaxy presented here would have an oxygen abundance of about 12+log(O/H)$\sim 7.05$. Although the mass-metallicity regime explored in \citet{mannucci10} is different, it appears that this galaxy does not seem to fit in a scenario of no evolution in the $M^{\star}$-$Z$-SFR fundamental plane. A large sample of galaxies is needed to test this scenario, and will be the subject of a forthcoming paper (Henry et al. in prep).  

\section{Summary}
\label{sec:summary}

The WISP survey \citep{atek10} offers a unique opportunity to search for high-redshift strong emission-line galaxies, down to a line flux limit of $\sim 5 \times 10^{-17}$ \ergscm, regardless of their continuum brightness, thanks to unprecedented IR grisms capabilities of the new WFC3 instrument onboard the {\em HST}.

We selected in this work a sample of 176 objects in 54 fields, with rest-frame EW higher than 200 \AA, spanning a redshift range of $0.35 < z < 2.3$. This selection results in a surface density of 1 object per square arcmin.

The presence of such strong emission lines in the spectra of star-forming galaxies at $0.8 <z<2.35$ has important implications for the study of high-redshift galaxy population. We first show that the contribution of nebular lines to the total broadband flux density can be more than 1 magnitude, with a median value at 0.3 mag. We demonstrate that strong emission lines falling in the IR broadband filters can mimic the $Y_{098} - J_{125}$ and $J_{125} - H_{160}$ color criteria used to select $z \sim 8$ galaxy candidates. While, the presence of such interlopers in deep HUDF observations is unlikely, the wide WFC3 pure parallel surveys are prone to such contamination because of the lack of deep optical observations.

High-EW emission lines can significantly affect the SED modeling of high-z galaxies and consequently the derived physical properties. We show that, when the emission lines create the appearance of a Balmer/4000 \AA\ break, the age and mass can be overestimated by an average factor of 2 and up to a factor of 10. More importantly, the situation will be much more problematic at higher redshift where we expect higher equivalent widths. Our results confirm empirically the theoretical predictions of \citet{schaerer09} in a large sample of galaxies and call for a careful treatment of nebular lines in SED fitting of star-forming galaxies at all redshifts.

The high-EW sample consists of young, unevolved galaxies with high specific star formation rates, which appear to be extreme outliers of the main sequence of the SFR-$M^{\star}$ relationship. The high-EW selection can sample metal-deficient galaxies. As an example, we obtained rest-frame optical spectroscopy with LRIS at {\em Keck} for 5 of these objects, but only one of them was at the right redshift for the metallicity measurement. This galaxy is located at $z \sim 0.7$ and is an extremely metal-poor galaxy. It follows the metallicity-luminosity relation derived from the USEL sample \citep{hu09} but is also consistent with $z \sim 0.7$ galaxies \citep{savaglio05}. With an oxygen abundance of 12+Log(O/H)$ = 7.47 \pm 0.11$, it is amongst the lowest metallicities in the samples of XMPGs \citep{kunth00,kakazu07,kewley07}. A larger sample of low-metallicity galaxies will be analyzed in a forthcoming paper (Henry et al. in prep), but we already demonstrate that we can find faint metal-poor galaxies in the high-EW sample to be targeted in follow-up spectroscopy.

\acknowledgments
We thank the anonymous referee for very useful comments that improved the content and the clarity of the paper, and Ranga-Ram Chary and Yuko Kakazu for interesting discussions. We thank Martin K\"ummel, Harald Kuntschner, Jeremy Walsh, Howard Bushouse, and the staff members of the Space
Telescope Institute for their help with the data reduction. We also wish to recognize and acknowledge the very significant cultural role and reverence that the summit of Mauna Kea has always had within the indigenous Hawaiian community.  We are most fortunate to have the opportunity to conduct observations from this mountain.
\bibliographystyle{apj}
\bibliography{ew}

\end{document}